\documentclass[%
 reprint,
superscriptaddress,
nofootinbib,
 amsmath,amssymb,
 aps,
floatfix,
]{revtex4-2}

\usepackage{graphicx}
\usepackage{float}
\usepackage{dcolumn}
\usepackage{bm}
\usepackage{siunitx}
\usepackage{xcolor}
\usepackage{amsmath}
\usepackage{hyperref}
\usepackage{lineno}
\usepackage[autostyle]{csquotes}
   \MakeAutoQuote{‘}{’}
\newcommand{\comment}[1]{}

\hypersetup{colorlinks,linkcolor={magenta},citecolor={magenta},urlcolor={magenta}}  

\usepackage{amsmath}
\DeclareMathOperator{\Tr}{Tr}
\begin{document}

\preprint{APS/123-QED}

\title{Probing early structure and model-independent neutrino mass with high-redshift CMB lensing mass maps}

\renewcommand{\vec}[1]{\mathbf{#1}}
\newcommand{\fr}[1]{\textcolor{blue}{(FQ: #1)}}
\newcommand{\tn}[1]{\textcolor{magenta}{(TN: #1)}}
\newcommand{\omar}[1]{\textcolor{green}{(OM: #1)}}
\newcommand{\mat}[1]{\textcolor{red}{(MM: #1)}}
\newcommand{\bds}[1]{\textcolor{cyan}{(BDS: #1)}}

\author{Frank J. Qu}
\email{jq247@cam.ac.uk}
\affiliation{DAMTP, Centre for Mathematical Sciences, Wilberforce Road, Cambridge CB3 0WA, UK}
\affiliation{Kavli Institute for Cosmology Cambridge, Madingley Road, Cambridge, CB3 0HA, UK}

\author{Blake D.\ Sherwin}
\affiliation{DAMTP, Centre for Mathematical Sciences, Wilberforce Road, Cambridge CB3 0WA, UK}
\affiliation{Kavli Institute for Cosmology Cambridge, Madingley Road, Cambridge, CB3 0HA, UK}

\author{Omar Darwish}
\affiliation{Universit\'{e} de Gen\`{e}ve, D\'{e}partement de Physique Th\'{e}orique and Centre for Astroparticle Physics,
24 quai Ernest-Ansermet, CH-1211 Gen\`{e}ve 4, Switzerland}
\affiliation{DAMTP, Centre for Mathematical Sciences, Wilberforce Road, Cambridge CB3 0WA, UK}

\author{Toshiya Namikawa}
\affiliation{Kavli Institute for the Physics and Mathematics of the Universe (WPI), UTIAS}

\author{Mathew S. Madhavacheril}
\affiliation{Department of Physics and Astronomy, University of Pennsylvania, 209 South 33rd Street, Philadelphia, PA 19104, USA}
\affiliation{Perimeter Institute for Theoretical Physics,
31 Caroline Street N, Waterloo ON N2L 2Y5 Canada}
\date{\today}

\begin{abstract}
CMB lensing maps probe the mass distribution in projection out to high redshifts,  but significant sensitivity to low-redshift structure remains. In this paper we discuss a method to remove the low-redshift contributions from CMB lensing mass maps by subtracting suitably scaled galaxy density maps, nulling the low redshift structure with a model-insensitive procedure that is similar to delensing. This results in a high-$z$-only mass map that can provide a probe of structure growth at uniquely high redshifts: if systematics can be controlled, we forecast that CMB-S4 lensing combined with a Rubin-LSST-like galaxy survey can probe the amplitude of structure at redshifts $z>3.75$ ($z>5$) to within $2.3\%$ ($3.3\%$). We then discuss other example applications of such high-$z$ CMB lensing maps. In standard analyses of CMB lensing, assuming the wrong dark energy model (or wrong model parametrization) can lead to biases in neutrino mass constraints. In contrast, we show with forecasts that a high-$z$ mass map constructed from CMB-S4 lensing and LSST galaxies can provide a nearly model-independent neutrino mass constraint, with only negligible sensitivity to the presence of non-standard dark energy models, irrespective of their parametrization.


\end{abstract}

\maketitle


\section{\label{sec:level1}Introduction}

As the Cosmic Microwave Background (CMB) photons travel from the surface of last scattering to our telescopes, they are deflected by the gravitational influence of matter in our Universe (See Ref. \cite{LEWIS_2006} for a review). This lensing effect leads to a remapping, by typically a few arcminutes, of the observed CMB anisotropies on the sky. The lensing is sensitive to structures over a  broad range of redshifts: most of the signal arises from between redshifts $z=0.5$ and $z=5$, although structures at higher redshifts also contribute to the signal; CMB lensing is hence a unique direct probe of the mass distribution at redshifts above $z \sim 4$. However, our ability to constrain these still unexplored high-redshift contributions with CMB lensing alone is limited by degeneracy with uncertainties in low-$z$ structure growth and geometry (e.g., uncertainties in the dark energy model) as well as sample variance from low-redshift structures. To accurately and precisely probe high-redshift structure, and for several other applications, it can hence be useful to isolate only the high-redshift contributions to the CMB lensing data. To achieve this, in this paper we propose to clean the low redshift contributions from the lensing field by subtracting suitably-scaled correlated tracers such as galaxy density (or galaxy lensing) maps. Our method for nulling the low-$z$ structure uses a process similar to delensing and similar to related approaches proposed by \cite{McCarthy2020}, with the difference that we do not account for noise in the process since our goal is to null the signal; this leads to more effective cleaning. (Related ``nulling'' ideas have also been presented in  \cite{Maniyar2021} and in the galaxy lensing literature \cite{Huterer_2005,Bernardeau_2014,Barthelemy_2020, Joachimi_2008,Heavens_2011}; our work focuses on galaxy density maps rather than intensity mapping lensing and does not assume perfect knowledge of the galaxy kernel to perform the cleaning.) We forecast how well such a high-$z$ mass map and lensing spectrum can be measured with upcoming surveys; we also discuss why our methodology is only weakly sensitive to the detailed properties (e.g., galaxy bias) of the galaxy tracer used for cleaning. 

Our method should be contrasted with cross-correlation-based approaches (e.g., \cite{Mishra-Sharma2018} involving CMB lensing, cosmic shear, and galaxies, \cite{2018Schmittfull} involving CMB lensing and galaxies), where a joint likelihood is constructed for all auto- and cross-spectra of galaxy survey observables together with CMB lensing. While such approaches can also provide powerful and indeed optimal constraints on high-$z$ structure, the constraints obtained generally involve also modelling the behaviour of low-$z$ structure (or at least the shape of low-$z$ spectra if a free amplitude is marginalized over), whereas our approach explicitly nulls all low-$z$ contributions at the cost of potentially increased errors. For example, if we wish to constrain neutrino masses using a standard cross- and auto-spectrum analysis despite uncertainties in the properties of dark energy, a model must be constructed for the low-redshift dark energy behaviour and the constraints may be biased if the model is incorrect or incomplete; in contrast, our cleaning procedure reduces our sensitivity to such model biases insofar as they appear only in the redshifts that are removed. Our method also has the advantage that the high-$z$ mass map, once constructed, can be easily used to perform a variety of high-$z$ cross-correlation and other analyses. This is analogous to the construction of foreground-cleaned ILC maps in CMB analyses \cite{Delabrouille_2008,Eriksen_2004,Abylkairov_2021}, where using a cleaned map is often more convenient and robust than performing a full joint analysis of spectra at all frequencies. \footnote{A toy picture of the difference between cross-correlation tomography and redshift-cleaning is as follows: imagine that we directly observed the 3d matter power spectrum in thin redshift shells $P(k,z_i)$ for 3d wave-number $k$ and redshift bin $z_i$. Cross-correlation tomography would be equivalent to constructing a likelihood for the entire data set (including the low redshifts), whereas our approach is akin to simply throwing out the low-redshift bins and only modelling the remainder. Of course, in practice, projection, galaxy bias and other effects make the real situation slightly more involved.}

As an example application of our high-$z$ lensing map (as previously mentioned), we will discuss in detail how one can use it to determine the unknown neutrino mass sum with reduced model dependence. The sum of the masses of the three neutrino species, $\sum{m}_\nu$, is a key cosmological observable that can be determined via the suppression of the CMB lensing signal it produces. While a cosmological detection of the sum of neutrino masses is expected within the next decade with high-resolution CMB experiments such as the Simons Observatory (SO) \cite{Ade_2019}, SPT-3G \cite{2014}, and CMB-S4 \cite{abazajian2019cmbs4}, these constraints are generally derived assuming a standard $\Lambda \rm CDM$ cosmological model. Neutrino masses can also be determined in extended models with more complex, $w_0-w_a$ dark energy behaviour, especially if galaxy survey data are included \cite{Allison2015, Mishra-Sharma2018, Yu2018,2019nu}, albeit with somewhat degraded constraints; however, there is always some degree of model-dependence even in fixing an appropriate parametrization of dark energy models, given that the physics of dark energy (or modified gravity leading to similar phenomenology) is still not fully known. 
 In contrast, we discuss in this paper how neutrino mass can be determined from the power spectrum of the aforementioned high-$z$ lensing map without being biased by assuming standard (or fixed $w_0-w_a$) dark energy or Einstein gravity; in this case, we need only to assume that at sufficiently high redshifts the effects of dark energy or modified gravity can be neglected, with the matter component dominating the energy density. 

In Appendix \ref{HR}, we also discuss another application of partial delensing of a CMB lensing mass map to modify its effective redshift origin: reducing the error on cross-correlation of CMB lensing with tracers restricted to only a certain narrow redshift range. \footnote{In particular, motivated by the fact that the cleaning method introduced here is flexible and can isolate any redshift range when an appropriately correlated tracer is used, we explore briefly the prospects of using the cosmic infrared background (CIB) to delens the high redshift content of the lensing map. This is useful in the context of cross-correlation analyses \cite{Darwish2020,Krolewski2020,2018,2016}. The low redshift lensing map is expected to have a higher correlation with low redshift galaxy tracer and can lead to better constraints on cosmological parameters such as the linear bias $b_1$ and low redshift amplitude of structure $A_\text{low}$.}

Our paper is structured as follows. We first present our lensing cleaning technique in Section \ref{Sec. Cleaning}; in Section \ref{Sec. amp} we introduce the datasets used and the cleaning method applied in the context of amplitude of structure measurements. In Section \ref{sec. weight}, we explain, with forecasts of biases, why using a high-$z$-only lensing map is a promising approach for neutrino mass measurements with minimal dark energy model dependence. We conclude in Section \ref{sec.discussion}, outlining other potential applications of our cleaning method. Finally, Appendix \ref{appendixA} discusses how this cleaning method removes potential degeneracies in
the determination of neutrino mass sum with effects induced by models of modified gravity and Appendix \ref{HR} explores the cleaning method in the context of improving the signal-to-noise ratio of lensing cross-correlations with galaxy fields.

Throughout this paper, unless stated otherwise, we assume a $\Lambda$CDM cosmology in a flat universe with fiducial parameters   $100\theta_{\text{MC}}=1.0409,\Omega_bh^2=0.0223,\Omega_ch^2=0.1198,\tau=0.06,n_s=0.9645,A_s=2.2\times10^{-9},\sum{m}_\nu=0.06\si{eV}$, which give the values of the approximated acoustic angular scale at recombination, the physical baryon density, the physical cold matter density, the optical depth at recombination, the slope and amplitude of primordial scalar fluctuation at a pivot scale of 0.05 $\si{Mpc^{-1}}$ and the neutrino mass sum, respectively. 

\section{Obtaining a high redshift lensing map}\label{Sec. Cleaning}

\subsection{Basic methodology}
In practice, we do not have direct access to high-redshift-only lensing field measurements; galaxy lensing surveys are generally restricted to $z<2-3$ due to challenges in observing well-characterised source galaxies in large numbers at high redshifts, and CMB lensing measurements are sensitive to a projection over a wide range of redshifts. However, introducing an external tracer field $\hat{X}$ which is correlated with the CMB lensing field at low redshifts can allow us to remove the low-$z$ portion of the CMB lensing field. The external mass tracer can either be a galaxy density field or a galaxy lensing map; although our methods are also applicable for galaxy lensing maps, we will consider the Rubin Observatory -- Legacy Survey of Space and Time (LSST) \cite{LSST}  galaxy density maps as a representative example, assuming these can be approximated as linearly biased on the large scales of interest.

How can we best remove the low-$z$ portion of the CMB lensing field using external large-scale structure tracers? We note that a similar, but slightly different goal was studied in analyses of delensing using large-scale structure (LSS) tracers \cite{Sherwin2015,2022SO,2017yu}. For B-mode delensing, the aim is to minimize the variance of the total field after a suitably-filtered LSS tracer has been subtracted; whether this variance arises from lensing signal or noise is irrelevant to the computation. In contrast, our goal in obtaining a high-$z$-only lensing map is only to remove the low-$z$ \emph{signal}; the effect of the (we assume, well-understood) noise is irrelevant for this cleaning. However, the formalism that must be applied is very similar. To clean out the low-redshift signal, we can simply apply standard LSS delensing methodology, with the important difference that we set noise to zero in our filtering, since we care primarily about minimizing signal power rather than minimizing the total power including noise.\\

Therefore, drawing on this simple modification of the previous delensing study \cite{Sherwin2015}, we can write down expressions that will give the best performance in removing the low-$z$ portion of the CMB lensing field. If we have a single tracer field, the lensing cleaning proceeds as follows:
\begin{equation}\label{eq. kk}
    \hat{\kappa}^{\text{clean}}_\vec{L}=\hat{\kappa}_\vec{L}-\frac{C^{{\kappa}{{X}}}_L}{C^{{X}{X}}_L}\hat{X}_{\vec{L}},
\end{equation}
where $\hat{\kappa}_\vec{L}$ is the original CMB lensing convergence, and we adopt the convention that quantities with/without a hat represents quantities with/without noise, i.e. $C^{\hat{X}\hat{Y}}_L=C^{{X}{Y}}_L+N^{{X}{Y}}_L$. This expression can be simply derived by minimizing the power spectrum of the linear combination ${\kappa}_\vec{L}-c(\vec{L}) {X}_{\vec{L}}$ with respect to the cleaning coefficient $c$, assuming no noise in the maps.

To see how the above accomplishes our goal in the removal of the low redshift portion of the lensing field let us decompose $\kappa = \kappa_{\text{low}} +\kappa_{\text{high}}$ into two uncorrelated low and high redshift pieces and assume, for illustration, that the galaxy field $X$ is perfectly correlated with the low redshift part via a scaling function $T_L$, $X_{\bm{L}}=T_L\kappa_{\text{low},\bm{L}}$.

The power spectrum of the cleaned lensing field, assuming noiseless $\hat{\kappa}_{\bm{L}}$ and $\hat{X}_{\bm{L}}$ fields is given by
\begin{equation}\label{eq.c}
    C^{\kappa_{\text{clean}}\kappa_{\text{clean}}}_L=C^{\kappa_{\text{high}}\kappa_{\text{high}}}_L+C^{\kappa_{\text{low}}\kappa_{\text{low}}}_L-\frac{(T_LC^{\kappa_{\text{low}}\kappa_{\text{low}}}_L)^2}{T^2_LC^{\kappa_{\text{low}}\kappa_{\text{low}}}_L}
\end{equation}
 We see that the last two terms of Eq. \ref{eq.c} cancel out, leading to the perfect subtraction of the low-$z$ contribution of the lensing power spectrum.  In practice, one does not expect the external tracer signal to be perfectly correlated with the CMB lensing signal arising from a certain low redshift range, but, especially if LSS tracers in several narrow redshift bins are available, a sufficiently high signal correlation can still be achieved, enabling the subtraction of much of the unwanted contribution to the lensing map.

To maximise the subtraction of the low-$z$ contribution to the CMB lensing kernel, we can combine different galaxy redshift bins to form an effective tracer map:
\begin{equation}\label{gal}
   \hat{X}_{\bm{L}} = \sum_i{c_{i,L}}{\hat{g}_{i,\bm{L}}}
   \,, 
\end{equation}
where  ${g_i}$ is the galaxy field of bin $i$. Following the methods in \cite{Sherwin2015}, but with the key difference that in the weights we only include signal power without (shot) noise, the coefficients $c_i$ which maximise the correlation coefficient between the signal part of ${X}$ and the lensing field are given by \begin{equation}\label{weights}
    c_{i,L}=\sum_j{(\text{Cov}^{gg}_L})^{-1}_{ij}C^{\kappa{g_j}}_L.
\end{equation}
Here, $\text{Cov}^{gg}_{L,ij}$ is the element of the covariance matrix between galaxies in bin $i$ and bin $j$. (Note that if galaxy bins are combined using the weights of Eq. \ref{weights}, this automatically implies that  the ratio $C^{\kappa{X}}_L/C^{XX}_L=1$.)
Although all spectra used to construct the weights can, in principle, be obtained empirically (or at least with a fit of a model to data), for forecasting it is worth having fully analytical expressions for these spectra. The signal part of the galaxy spectra is then given explicitly in the Limber approximation \cite{limber} as 
\begin{equation}
   \text{Cov}^{gg}_{L,ij} =\int\frac{dzH(z)}{\chi^2(z)}W^i(z)W^j(z)P\Big(k=\frac{L+\frac{1}{2}}{\chi(z)},z\Big) .
\end{equation}
Here $H(z)$ is the Hubble parameter, $\chi (z)$ is the comoving angular diameter distance to redshift $z$ and $P(k,z)$ is the matter power spectrum with wavenumber $k$ and redshift $z$. For the $i$th galaxy bin, the window function is given by
\begin{equation}
    W^{i}(z)=\frac{b_i(z)dn_i/dz}{\int{dz^\prime}dn_i/dz^\prime}.
\end{equation}
The bias $b_i$ and the redshift distribution $dn_i/dz$ used in our forecasts are specified in detail in section \ref{sec. data} below.  

As indicated above, the difference with the approach introduced for delensing is that we are interested in nulling the CMB signal in the cleaning case, and variance minimization is not the primary concern. Therefore, we emphasize again that we do not include galaxy shot noise  $N_L^{g_ig_j}=\delta_{ij}/{n_i}$ in the above covariance matrix.

Finally, the lensing auto/cross spectra are computed in a similar manner as
\begin{equation}\label{eq. generic_cl}
   C^{\alpha\beta}_{L} =\int\frac{dzH(z)}{\chi^2(z)}W^\alpha(z)W^\beta(z)P\Big(k=\frac{L+\frac{1}{2}}{\chi(z)},z\Big),
\end{equation}

where $\alpha,\beta\in(\kappa,g_i)$.

For CMB lensing, the convergence  kernel $W^\kappa(z)$ is given by
\begin{equation}
    W^\kappa(z)=\frac{3}{2H(z)}\Omega_mH^2_0(1+z)\chi(z)\Bigg(\frac{\chi_\star-\chi{(z)}}{\chi_\star}\Bigg)
\end{equation}
where $H_0$ and $\Omega_m$ are the Hubble parameter and matter density today, respectively, and $\chi_\star$ is the comoving distance to the last scattering surface. 
The publicly available Boltzmann code \texttt{CAMB} \cite{Lewis_2000} was used to calculate the above auto- and cross spectra.

\subsection{Potential limitations: Uncertainty in the galaxy survey properties}

In the cleaning method above, an important step is the calculation of the weights $c_i$ in Eq. \ref{weights} used to maximize the correlation between the galaxy fields and lensing. When doing so, a model is assumed for the lensing cross spectra $C^{\kappa{g_i}}_L$ and auto spectra $C^{{ g_i g_i}}_L$  of the galaxy fields. Although good measurements are achievable for the LSST survey, the true spectra, in particular, the true cross spectra $C^{\kappa{g_i}}_L$ between the galaxy fields and CMB lensing are not known precisely. The fact that we do not know the true galaxy cross- and auto spectra is a potential problem for two reasons: first, a misestimation of the weights can lead to a wrongly weighted galaxy field $\hat{X}$ and hence give a residual $C^{\kappa_{\text{clean}}\kappa_{\text{clean}}}_L$ which is insufficiently (suboptimally) cleaned of low-$z$ contributions; second, even if the weights are chosen optimally, our ignorance of the spectra and other properties of the galaxy tracer such as biases and redshift distributions implies that the interpretation of the cleaned lensing signal (including, e.g.,~its redshift distribution) is complex. 

We will begin by discussing the first challenge: to what extent does uncertainty in the weights lead to suboptimal cleaning performance? We will show that the effects of fluctuations in the lensing-galaxy cross spectra result only in sub-percent changes in the cleaned spectra, which are negligible for our example analysis. We do this by generating 1000 Gaussian realisations of the cross-correlation spectra $C^{\kappa{g_i}}_L$ curves,  all of which are consistent with the forecast CMB-S4 and LSST errors (see the data section Sec \ref{data}) , and use these to construct perturbed weights $\tilde c_i$ of Eq. \ref{weights}.  From these weights we obtain 1000 realisations of $C^{\kappa_{\text{clean}}\kappa_{\text{clean}}}_L$  which can then be compared with  the fiducial one, $C^{\kappa_{\text{clean}}\kappa_{\text{clean}},\text{fid}}_L$ . 

\begin{figure}
\includegraphics[width=\linewidth]{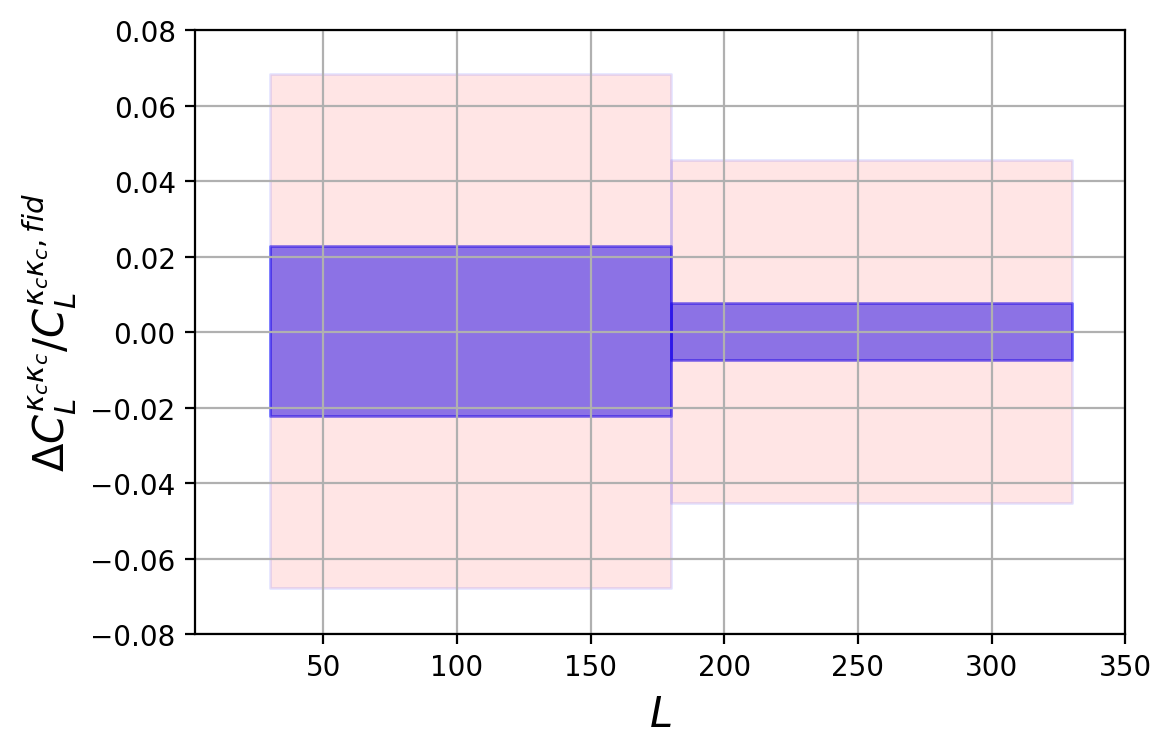}
    \caption{\label{fig:fid_diff} Standard deviation of the cleaned lensing power spectra constructed from different realisations of the weights used to clean out the low redshifts (blue); the weights are, in turn, obtained from different realisations of ${C^{\kappa{g_i}}_L}$ measurements in two broad multipole $L$ bins (obtained by perturbing the fiducial model by a realistic amount). It can be seen that the uncertainty due to the weights is negligible, as it is only a small fraction of the total uncertainty of the cleaned lensing spectra (red).}
\end{figure}

The mean fractional difference between the cleaned power spectrum obtained from these perturbed weights and that of the fiducial cleaned spectra is shown in Fig. \ref{fig:fid_diff}, where we have applied a broad multipole binning with edges at $(40,190,340)$. We see that this difference is close to zero, with a standard deviation of 0.02 and 0.007 in the first and second bin respectively. These correspond to only $32\%$ and $16\%$ of the total uncertainties of the cleaned lensing spectra. 

Why are the deviations in the cleaned, high-redshift lensing spectra small? We will attempt to explain this by computing the effect that the uncertainty in the weights has in the final cleaned lensing spectrum. The perturbed cleaned lensing field is given by
\begin{equation}
    \hat{\kappa}^{\text{clean}}_\vec{L}=\hat{\kappa}_\vec{L}-    \sum_i\tilde c_i[C^{ab}_\ell(1+\epsilon_\ell^{ab})]g_i ,
\end{equation}

where $ab$ stands for every possible auto- and cross-spectrum of the LSS tracers and the CMB lensing field.

The weights $\tilde c_i$, which are a function of the true spectra $C^{ab}_\ell$, are chosen to minimize the power spectrum of the cleaned lensing field $C^{\kappa_{\text{clean}}\kappa_{\text{clean}}}_L$ .

Since the optimal weights are the result of a minimization of the cleaned signal power, the cleaned power has no linear sensitivity to small changes in the weights; this implies that the cleaned power also has no linear sensitivity to small changes $\epsilon^{ab}(\ell)$ in the fiducial spectra that are used to construct the weights.  It follows that the signal in the cleaned lensing spectra only has a second-order dependence on small inaccuracies in the fiducial spectra:

\begin{equation}
    C^{\kappa_{\text{clean}}\kappa_{\text{clean}}}_L=C^{\kappa_{\text{clean}}\kappa_{\text{clean}},\text{fid}}_L + \mathcal{O} ((\epsilon^{ab})^2).
\end{equation}

This explains the result we saw in Fig. \ref{fig:fid_diff}: even moderate uncertainties in the spectra used to weigh the tracers do not prevent the construction of a high-$z$ mass map, since fractional errors only enter quadratically. Therefore, for weighting, our cleaning procedure can be applied without requiring precise knowledge of the tracer redshift distribution (or other properties), relying only on observed spectra at modest precision. As previously mentioned, there is a good analogy between our cleaning method and the ILC methods \cite{Delabrouille_2008,Eriksen_2004} commonly used for foreground cleaning in CMB data analysis; in both cases, the cleaning can be performed using only observed spectra.

Of course, suboptimal weighting due to incorrectly assumed spectra is not the only concern when applying the cleaning procedure we have described. An additional complication is that, even if an optimal weighting is applied, there may be remaining uncertainty in modelling the cleaned spectra, because the bias and redshift distribution of the galaxy sample used for cleaning may be uncertain. Unfortunately, for redshift uncertainty, this is an obstacle that must be overcome by directly characterizing the redshift origin of the galaxies, as without knowledge of the galaxy redshift distribution the redshift origin of the cleaned field must remain uncertain as well. On the other hand, our cleaning procedure naturally accounts for the unknown bias, assuming that the bias is linear and simply re-scales the spectra in a narrow bin (within which we can assume a redshift evolution). To see this, we note that the cleaning tracer is given by

\begin{equation}
   {X}=\sum_i{c_i}{g_i} = \sum_i{c_i}{b_i \Delta_i}
\end{equation}
where we have divided out the linear bias $b_i$ from the projected galaxy density field to give the projected matter density fluctuation $\Delta_i$. Our expression for the weights $c_i=\sum_j{(\text{Cov}^{gg}})^{-1}_{ij}C^{\kappa{g_j}}_L$ implies that the resulting cleaning tracer (and hence the cleaned map) is independent of the linear bias, which can be seen as follows:

\begin{align}
   {X}=\sum_i{c_i}{g_i} &= \sum_{ij} {(\text{Cov}^{g_i g_j}})^{-1} C^{\kappa{g_j}}_L {b_i \Delta_i}\nonumber\\
   &= \sum_{ij} {\frac{1}{b_i b_j}(\text{Cov}^{\Delta_i \Delta_j}})^{-1} b_j C^{\kappa{\Delta_j}}_L {b_i \Delta_i}   \nonumber\\
   &= \sum_{ij} {(\text{Cov}^{\Delta_i \Delta_j}})^{-1} C^{\kappa{\Delta_j}}_L { \Delta_i},
\end{align}
where the bias has cancelled out. Of course, this is only true if our cleaning weights are correct; however, we note that by our previous argument, our results are insensitive to small errors in the weights, and our cleaning procedure should be able to derive the weights sufficiently well from the measured spectra.

\section{Measuring the amplitude of structure with high redshift lensing maps}\label{Sec. amp}

To illustrate our cleaning method, we will forecast the ability to use a high-$z$ only lensing map to probe the amplitude of the high-redshift structure. We first introduce the CMB-S4 lensing and LSST galaxy survey specifications we consider throughout this paper.

\subsection{Forecasting data used}\label{sec. data}

\subsubsection{Lensing specifications}\label{lens spec}
\begin{figure}
\includegraphics[width=\linewidth]{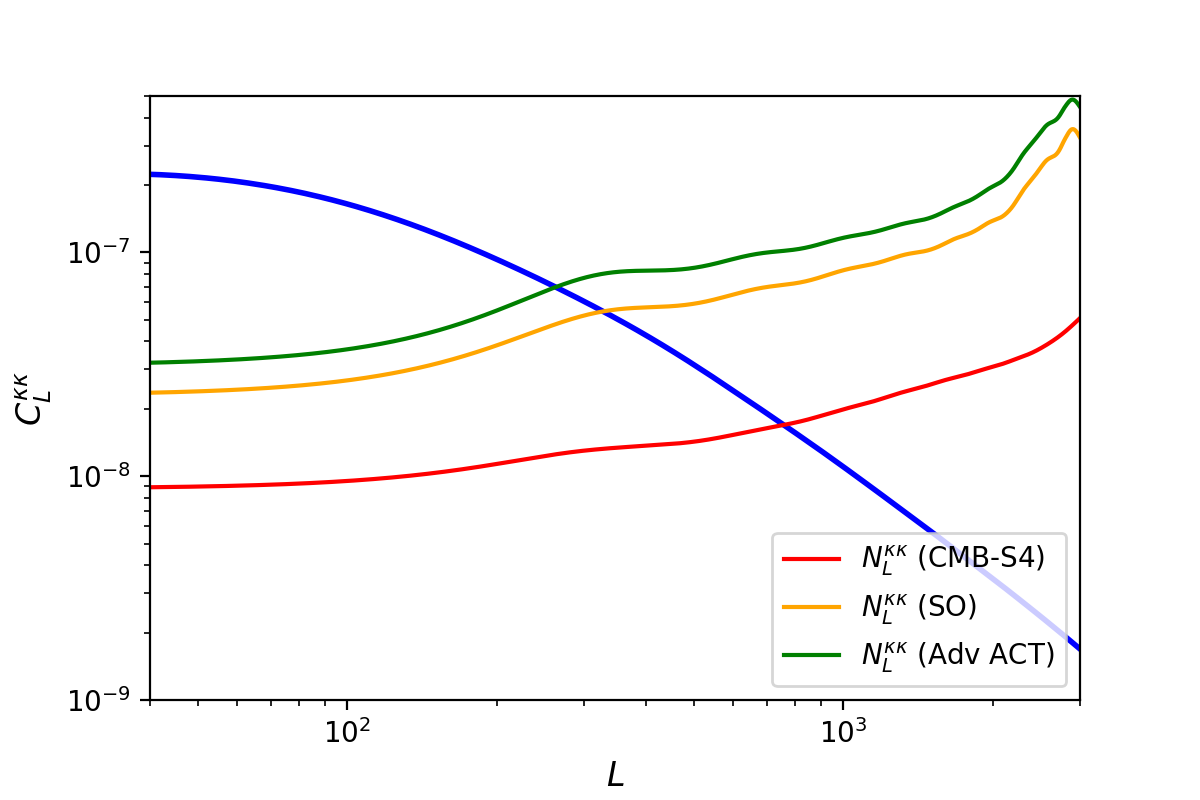}
\caption{\label{fig:lensing} CMB lensing power spectrum $C^{\kappa\kappa}_L$ (solid blue) and expected reconstruction noise $N^{\kappa\kappa}_L$ for an AdvACT-like, SO-like, and CMB-S4-like survey. The noise curves for AdvACT- and SO-like surveys were computed using \texttt{tempura} assuming quadratic estimator lensing reconstruction on the full sky with the specifications beam FWHM = $1.4'$, $\Delta_T=15\si{\mu{K}}'$, $\Delta_P=15.3\si{\mu{K}}'$,$f_{\text{sky}}=0.3$ and
beam FWHM = $1.4'$, $\Delta_T=8\si{\mu{K}}'$, $\Delta_P=11.3\si{\mu{K}}'$ and $f_{\text{sky}}=0.3$ respectively. The CMB-S4 noise arises from forecasts of iterative lensing reconstruction performance \cite{s4collaboration2020cmbs4}.}
\end{figure}

For CMB lensing, we use a CMB-S4-like \cite{s4collaboration2020cmbs4} experiment with the following specifications: beam FWHM = $1.4'$, noise levels $\Delta_T=1\si{\mu{K}}'$, $\Delta_P=1.4\si{\mu{K}}'$, and sky area $f_{\text{sky}}=0.4$. \comment{We compute the forecast mimimum variance reconstruction noise using the \texttt{tempura}\footnotetext{https://github.com/simonsobs/tempura} code}. We use lensing reconstruction noise curves assuming an optimal (``iterative'') measurement pipeline for CMB-S4, which is expected to have better performance than the standard quadratic estimator \cite{louis}. For temperature, we restrict the CMB multipoles to a range $50<\ell<3000$ to minimize foreground biases and increase this range to $\ell_{\text{max}}=5000$ for polarization, since polarization data are less contaminated by extra-galactic foregrounds. For Sec. \ref{sec. weight} we will also include information from the primary CMB temperature and polarization power spectra to break degeneracies between parameters \cite{1999eft}. The minimum variance reconstruction noise from combining the different temperature and polarization channels for the different CMB experiments is shown in Fig. \ref{fig:lensing}. 

\subsubsection{LSST specifications}
\begin{figure}
\includegraphics[width=\linewidth]{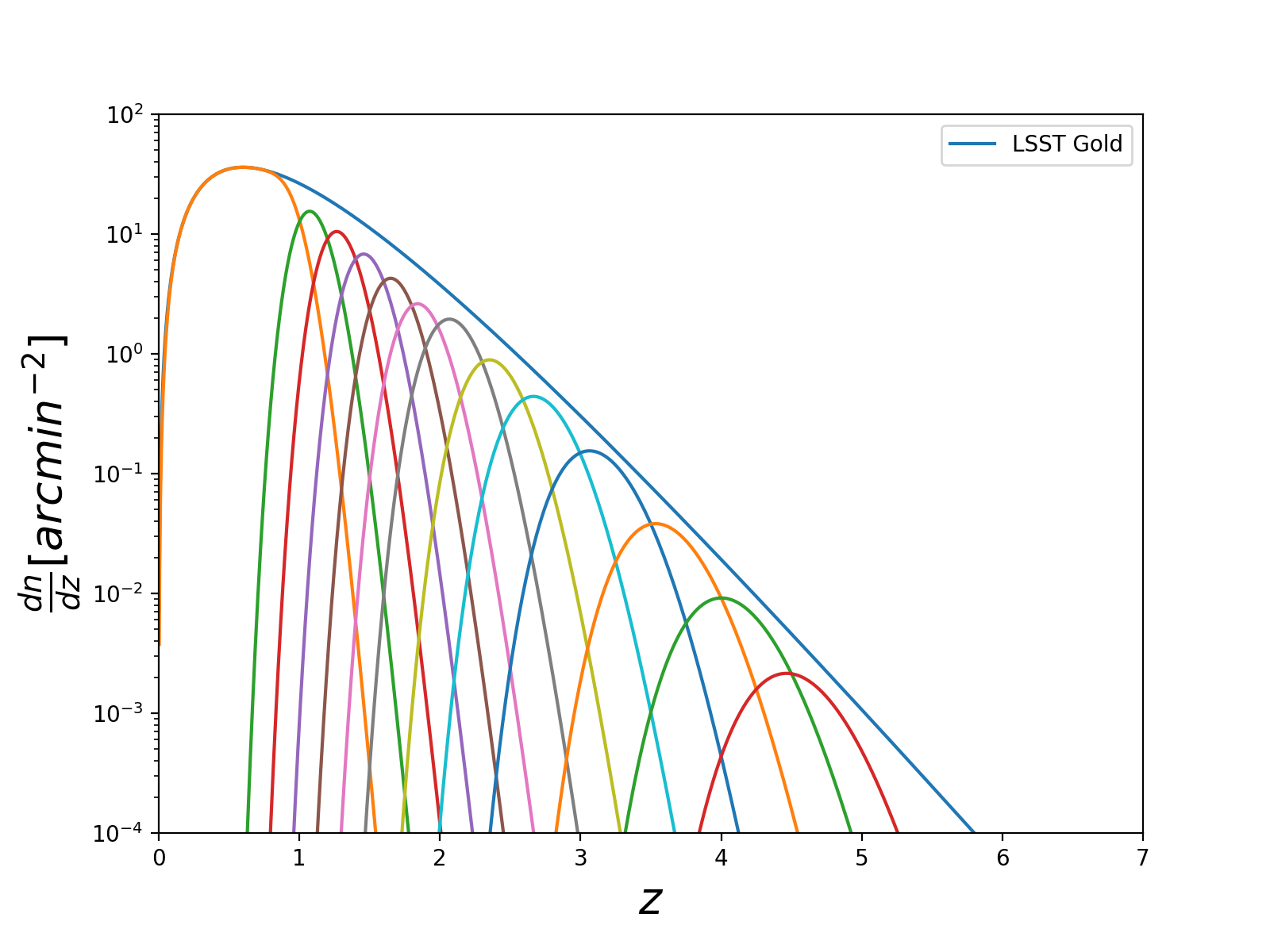}
\caption{\label{fig:lsst} The redshift distribution of the LSST Gold galaxy samples. We use 13  tomographic redshift bins in the range $0<z<5$. }
\end{figure}

For the Rubin-LSST galaxy survey, we use the Gold sample of galaxies with the following redshift distribution:

\begin{equation}
    \frac{dn}{dz}\propto\frac{1}{2z_0}\Big(\frac{z}{z_0}\Big)^2e^{-\frac{z}{z_0}},
\end{equation}
with $z_0=0.3$. This corresponds to $\bar{n}=40\si{arcmin}^{-2}$. We split the LSST kernel into 13  tomographic bins with bin edges $z=[0,1,1.2,1.4,1.6,1.8,2,2.3,2.6,3,3.5,4\allowbreak,4.5,5]$ . The above splitting provides enough freedom to rescale the galaxy kernels to match the profile of the CMB lensing kernel while still having a high signal to noise in each individual galaxy bin. For each bin $i$, we assume the linear galaxy bias is given by $b_i(z)=B_i(1+z)$, where $B_i$ is the overall bias amplitude with a fiducial value of $B_i=1$. To reduce the sensitivity of our forecasts to the uncertainties of non-linear modelling, we remove modes in the highly non-linear regime by setting a small scale cutoff in $k_{\text{max}}$ (for each redshift bin, $0.3h\si{Mpc^{-1}}$). A survey area of $18,000 \si{deg^2}$ corresponding to $f_{\text{sky}}\approx0.4$ is adopted; motivated by the planned survey regions, we assume that this has complete overlap with the CMB lensing survey. We partially account for photometric redshift errors by convolving the window function with the probability distribution function $p(z_{ph}|z)$ of the photometric redshift $z_{ph}$ at a given $z$ which is taken to be a Gaussian \cite{Chen2021}. The modified redshift distribution in each bin is then given by:
\begin{equation}
    \frac{dn_i}{dz}=\frac{1}{2}\frac{dn}{dz}\Big[\text{erf}\Big(\frac{z-z_i}{\sqrt{2}\sigma_z}\Big)-\text{erf}\Big(\frac{z-z_{i+1}}{\sqrt{2}\sigma_z}\Big)\Big]
\end{equation}
where $\sigma_z=0.05(1+z)$ is the width. Fig. \ref{fig:lsst} shows the LSST redshift distribution for the overall Gold sample and that of the 13 tomographic bins. 

\subsection{Measuring high redshift amplitude of structure}
\begin{figure}
\includegraphics[width=\linewidth]{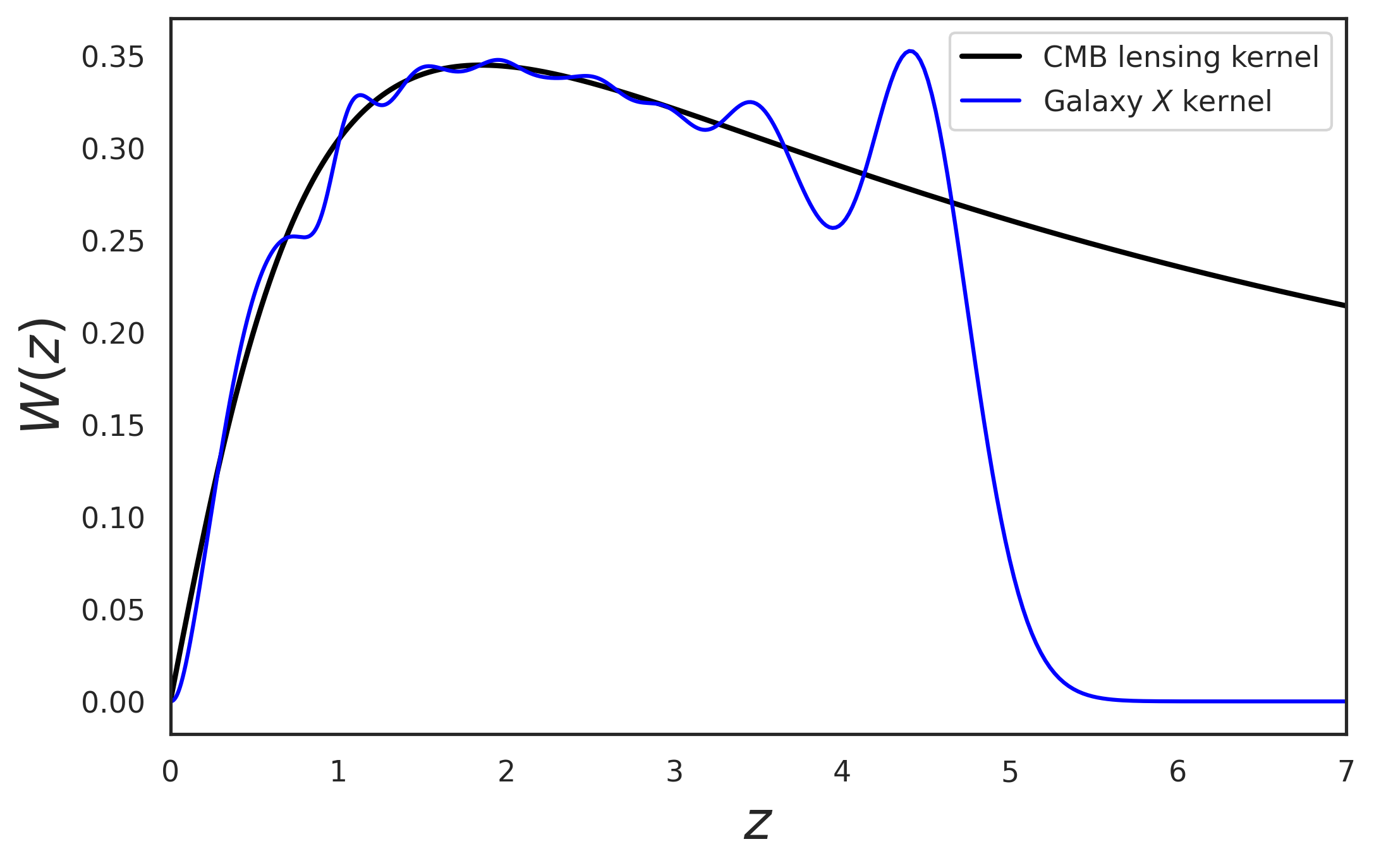}%
\caption{\label{fig. gal_window} Lensing window function in black and the effective galaxy window function obtained by combining the LSST galaxy bins up to $z \sim 5$ using weights calculated with Eq. \ref{weights} overlaid on top (blue).}
\end{figure}

We proceed to forecast the performance of high-$z$-only lensing maps obtained using the above LSST galaxies and following the procedure introduced in Sec. \ref{Sec. Cleaning} to delens the low redshift contribution to the lensing field.  Although the individual LSST galaxy window functions might not match well with the CMB lensing kernel, combining the samples using Eq. \ref{weights} results in an effective galaxy field ${X}=\sum_i{c_i}{g_i}$ that has a window function  $W^X=\sum_i{c_i}{W^{g_i}}$  which closely follows the CMB lensing kernel, as illustrated in Fig. \ref{fig. gal_window}.

\begin{figure}
\includegraphics[width=\linewidth]{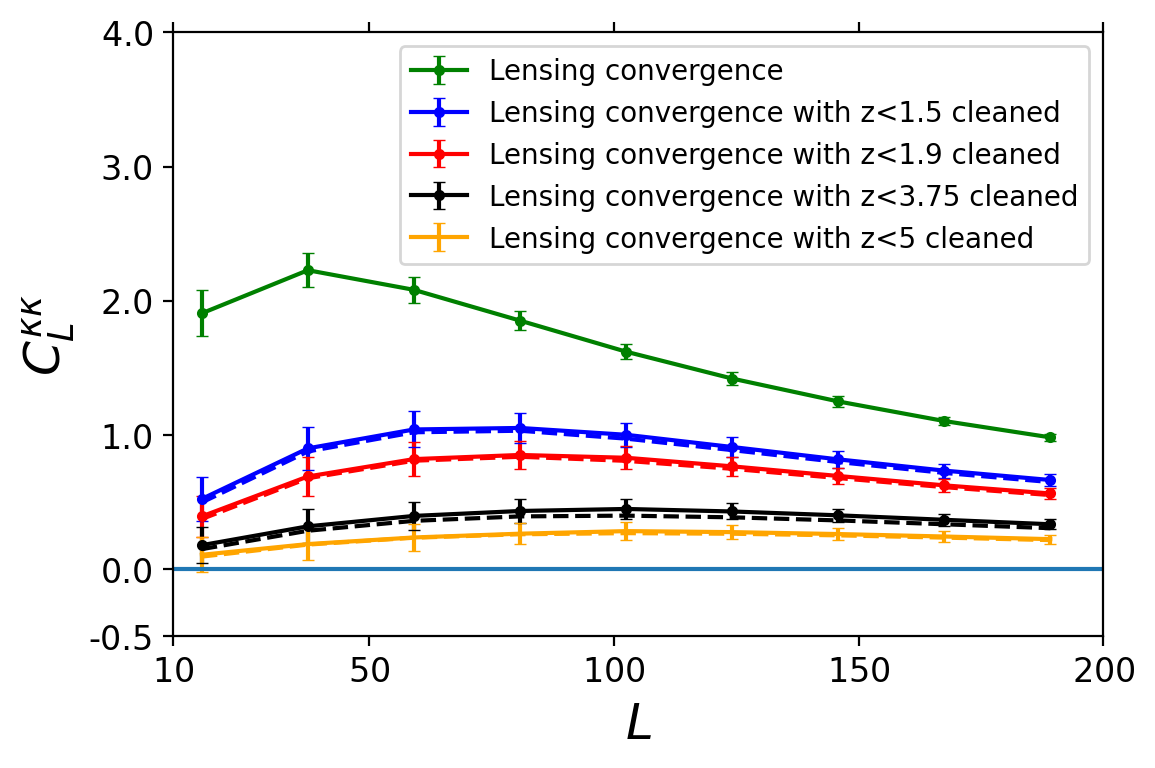}%
\caption{\label{fig. kappakappa} Forecast CMB lensing power spectra with low redshifts cleaned; the different colored curves show the results of cleaning with LSST galaxy density maps extending only to a certain redshift $z$. The fiducial uncleaned lensing power spectrum (green) is shown as a reference. As a comparison, the dashed line corresponds to the lensing spectra obtained using the lensing window function with the low redshift contribution perfectly removed.}
\end{figure}

These high-$z$ only lensing maps translate into high-$z$ lensing spectra from redshifts ($z>$ 0,~1.5,~1.9,~3.75 and 5) that can be determined (assuming tracers can be used up to wavenumbers of $0.3h\si{Mpc^{-1}}$, which translates into cleaned lensing spectra with $L_\text{max}=200$) at a signal-to-noise ratio of (83,33,29,18,13). The different lensing spectra are shown in Fig. \ref{fig. kappakappa}; the spectra of the cleaned lensing fields can be compared with the spectrum of the full lensing field in green. As expected, the removal of the low-$z$ structure leads to spectra with a lower amplitude and with the peak shifted to smaller scales (due to lensing occurring at greater distances.)

\begin{figure*}
\includegraphics[width=\linewidth]{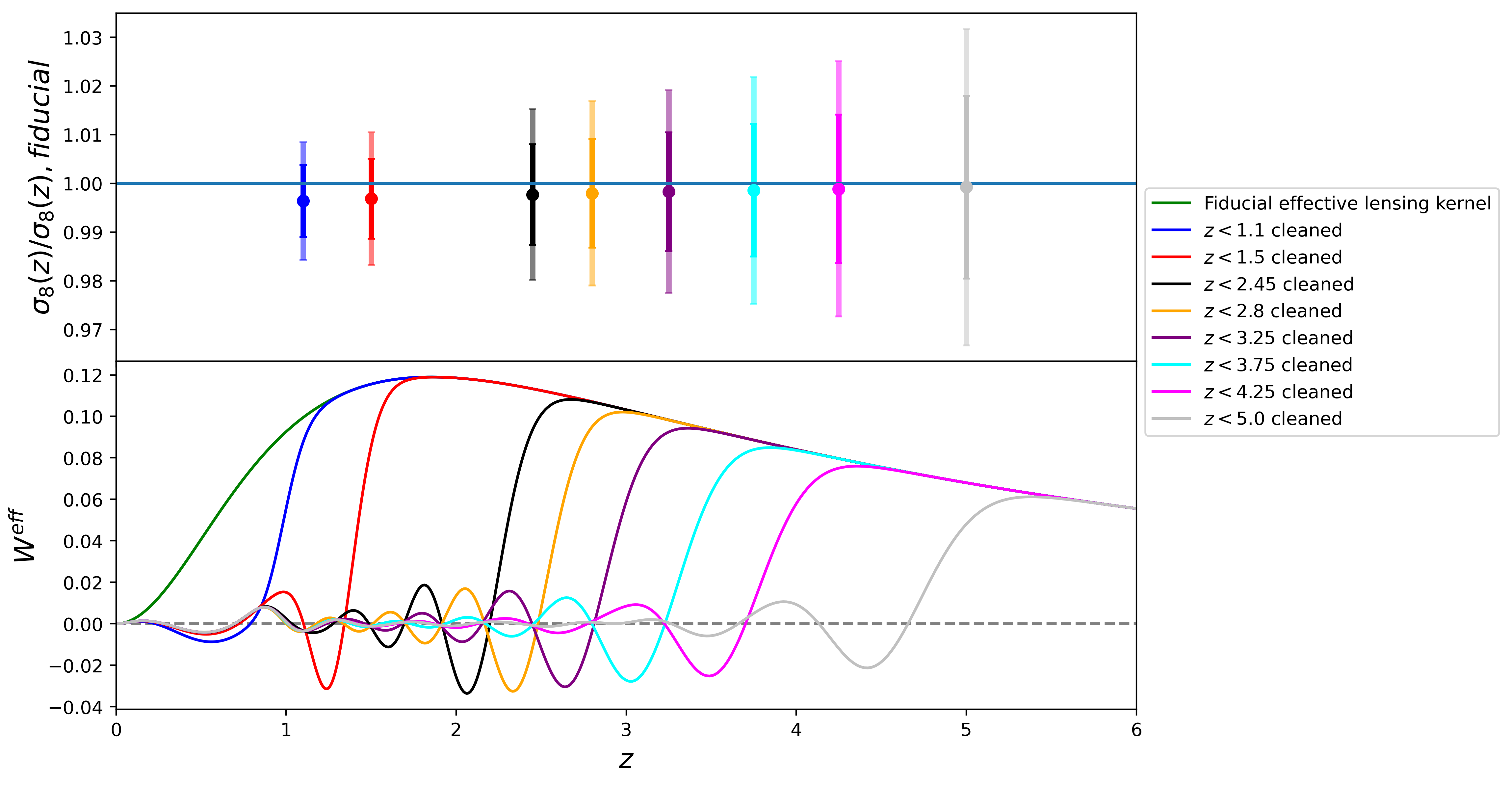}
    \caption{\label{Fig. sigma8} \textit{Top panel}: 1$\sigma$ constraints on $\sigma_8$ at high redshifts after cleaning the low-redshift contribution to CMB-S4 lensing maps using LSST galaxies. Each measurement is placed at a redshift where the effective lensing kernel peaks (see bottom panel). The smaller/larger errors corresponds to using lensing $L_\text{max}=400$ and $L_\text{max}=200$ respectively.
    \textit{Bottom Panel}: Effective lensing window function $W^{\text{eff}}$ for the fiducial uncleaned lensing in green and for lensing fields cleaned using LSST galaxies. 
}
\end{figure*}

The reduction in sensitivity of these cleaned lensing fields to low-$z$ structures enables a tomographic measurement of the amplitude of structure, $\sigma_8$ at multiple redshifts, which can be determined at a precision given by $\sim2$ times the lensing signal-to-noise ratio. This follows from the fact that each lensing field is sensitive to the amplitude of structure $\sigma_8$, and hence the lensing power spectrum $C^{\kappa\kappa}_L$ is directly proportional to $\sigma^2_8$, assuming all other parameters are held fixed. (Fixing other parameters appears to be a reasonable approximation for this forecast since we are only testing for several-percent-level departures from $\Lambda$CDM structure growth at high redshift, and other relevant parameters should be determined to much higher precision by the time the final CMB-S4 and LSST datasets are available.)

The top panel of Fig. \ref{Fig. sigma8} shows the determination of $\sigma_8$ as a function of the redshift up to which cleaning is performed.
Using lensing alone, one can probe the amplitude of structure at redshifts $(z=0,~1.5,~1.9,~3.75, 5)$ to within $\sim (0.6,1.4,1.5,2.3,3.3) \%$ respectively and $\sim (0.3,0.8,0.9,1.4,1.8) \%$ if using $L_{\text{max}}=400$ instead of $L_{\text{max}}=200$. CMB lensing alone hence can probe the growth history over a range of uniquely high redshifts where the $\Lambda${CDM} model has not yet been tested extensively.

We can further check that we are indeed insensitive to the low-$z$ information after cleaning by examining the effect on the lensing window function $W^\kappa$, which appears quadratically in the computation of the power spectra as described by Eq. \ref{eq. generic_cl}.
Denoting $W^{\text{eff}}(z)$ as the analogous term to $(W^{\kappa})^2$ appearing in the cleaned lensing spectra, one can see that the effect of cleaning is to reduce and shift the peak of this kernel to higher redshifts

\begin{equation}\label{eq. weff}
   W^{\text{eff}}(z)=(W^{\kappa})^2-W^XW^\kappa
\end{equation}

The above can be obtained  by looking at the window function appearing in the cleaned lensing spectra, which in the noiseless limit is given by
\begin{align}
    C^{\kappa\kappa,c}_L&=C^{\kappa\kappa}_L-C^{\kappa{X}}_L\nonumber\\&=\int\frac{dzH(z)}{\chi^2(z)}\Big((W^{\kappa})^2-W^XW^\kappa\Big){P}\Big(k=\frac{L+\frac{1}{2}}{\chi(z)},z\Big)\nonumber\\&=\int\frac{dzH(z)}{\chi^2(z)}W^{\text{eff}}(k)P\Big(k=\frac{L+\frac{1}{2}}{\chi(z)},z\Big)
\end{align}

The bottom panel of Fig. \ref{Fig. sigma8} shows the resulting $W^{\rm eff}$.  As seen already in Fig. \ref{fig. gal_window}, the galaxy window function $W^X$ follows closely to $W^\kappa$ at low redshifts, such that the effective window function in Eq. \ref{eq. weff} is close to zero in the redshift range where cleaning is applied and with the peak at $z=2$ shifted towards higher redshifts. The small oscillations around zero arise because the shapes of the galaxy windows functions do not match perfectly with that of the lensing kernel; furthermore, some overlap exists between the galaxy bins, which leads to over-or under-subtraction of the lensing window. Perfect nulling can be achieved in the ideal case when the galaxy bins are non-overlapping bins that are either much narrower than the distance over which the CMB lensing kernel varies or that have a redshift distribution that perfectly matches the CMB lensing kernel over the bin range.

\section{Weighing neutrinos with high redshift only lensing maps}\label{sec. weight}

A key goal of current cosmology and particle physics experiments is the measurement of the unknown mass of neutrinos \cite{2015neutrino}. Neutrinos initially behave as relativistic radiation in the early universe; as their temperature falls, they become non-relativistic, giving an energy density that evolves like Cold Dark Matter (CDM). 

A clear signature of massive neutrinos is a scale dependent suppression in the matter power spectrum, which can be understood as follows. On all scales, massive neutrinos contribute significantly to the total mean energy density of the universe, increasing its late-time expansion rate beyond the massless neutrino case; this increased expansion tends to suppress structure growth. However, on large scales, this is fully compensated by a corresponding increase in the strength of clustering and gravitational driving, causing massive neutrinos to produce the same structure growth as CDM. In contrast, on small scales, the growth suppression is not compensated because, due to the large thermal velocities that neutrinos have, they free-stream, erasing perturbations in the neutrino component.

Measurement of this suppression is the primary way in which the neutrino mass sum $\sum{m}_\nu$ can be measured with cosmological observations \cite{2012pastor,Allison2015}. This can be achieved by comparing the initial high redshift amplitude of structure obtained from the CMB power spectrum (via measurements of the optical depth) against a low redshift probe like CMB weak lensing \cite{2003knox,1998hu}. However, one caveat is that, because CMB lensing probes the total projected matter distribution down to $z=0$, it (along with nearly all other probes) is also sensitive to the growth suppression effects induced by dark energy. The suppression effect of neutrinos on the growth rate of the  overdensity $\delta_{\text{CDM}}$ during matter domination is described by $    \delta_{\text{CDM}}\propto[a]^{1-\frac{3}{5}f_\nu}
$  on scales much smaller than the neutrino free-streaming scale \cite{1998hu,2012pastor}, where $f_\nu$ is the fraction of matter-energy density in neutrinos, proportional to the neutrino mass.

Dark energy can similarly cause a suppression of the growth of matter structure, often described as follows: $\delta_{\text{CDM}}\propto[ag(a)]$, where $g(a)$ is a scale-independent damping factor (with a value $<1$ as the dark energy density becomes significant). Since both neutrino mass and dark energy result in a suppression of structure growth, non-standard dark energy physics could lead to a biased determination of the neutrino mass.

We will here consider applying our cleaning method to obtain a high-redshift-only lensing map, and use this to constrain $\sum{m}_\nu$. This is motivated by the fact that the damping factor due to dark energy $g(a)$ differs only from 1 when the dark energy density is non-negligible; hence, assuming that dark energy is only relevant below a certain low redshift, removing the lower redshift information means that the suppression effects in the matter power spectrum are due to neutrinos alone.

We expect the use of a high-$z$  lensing map to not greatly decrease the precision of a neutrino mass measurement for two reasons: first, upcoming measurements of neutrino mass are limited by our knowledge of the high-redshift amplitude of structure via the CMB optical depth, so that they are only weakly sensitive to a reduction in the precision of lower-redshift measurements; second, because the suppression signal is only logarithmic in $a$ or redshift so that the size of the effect is similar regardless of whether we measure the low-$z$ amplitude of structure at $z=0.5$ or $z=3$.\\

We will test this expectation quantitatively in the next section, performing forecasts to investigate whether a high-$z$ lensing map can give high-precision neutrino mass measurements with minimal dark energy model dependence. Before doing so, we can gain some intuition about why delensing the low redshift content helps decrease the neutrino mass bias by examining the lensing power spectrum Eq. \ref{eq. generic_cl} , written out again for clarity

\begin{equation}
    C^{\kappa\kappa}_L = \int^{z_\star}_0 dz \frac{H(z)}{\chi^2(z)}[W^{\kappa}(z)]^2P_{\delta\delta}\Big(k=\frac{L+\frac{1}{2}}{\chi(z)},z\Big)
    \,. 
\end{equation}

The presence of dark energy can affect the above in two ways:
\begin{enumerate}
\item The matter power spectrum $P_{\delta\delta}$ is obtained from the primordial scale independent power spectrum by applying the appropriate transfer function $T(k)$ (which accounts for the fact that growth is suppressed for modes which enter the horizon during radiation domination) and is also linearly proportional to a scale-independent damping factor $g(z)$ \cite{Pan2015}. This damping factor $g(z)$ describes the suppression of growth in the presence of dark energy (due to potential decays on all scales); it is normalised such that it is unity during matter domination.

\item Dark energy can also affect the geometry of the universe by changing the redshift dependence of the radial comoving distance, $\chi(z)\rightarrow{\chi^\prime(z)}$. We will argue in the following paragraphs that the effect of this on lensing is expected to be small at high redshifts once we fix the angular scale of the sound horizon; this scale is effectively fixed when we combine our lensing measurements with those of the primary CMB, which constrain the angle subtended by the sound horizon at very high precision. 
\end{enumerate}

By delensing the low redshift matter perturbations, we remove completely the effects that dark energy has on growth since there are no longer any affected matter perturbations contained in the high redshift lensing maps. The only possible effect that dark energy can have is  on geometry by altering the radial comoving distance to redshift $z$ given by

\begin{equation}
    \chi(z)=\int^z_0\frac{dz^\prime}{H(z^\prime)} .
\end{equation}

However, because the distance to recombination $\chi_\star=\chi(1100)$ is fixed to high accuracy by including CMB power spectrum data (via their measurement of the angular scale of the acoustic peaks) \cite{Aghanim2020}, in such an analysis $\chi(z)$ at high redshifts is well constrained and nearly unaffected by dark energy. One can see this by writing $\chi$ as

\begin{equation}
    \chi(z)=\chi_\star-\int^{z_\star}_z\frac{dz^\prime}{H(z^\prime)}
\end{equation}

The integral contains only $z>z_{\text{DE}}$ and the Hubble parameter at these high redshifts is unaffected by dark energy since (we assume) its energy density is negligible at these redshifts.

\subsection{Data Used and Forecasting method}\label{data}


\subsubsection{Experimental setup}

We employ the same CMB-S4 specifications for primary CMB and CMB lensing introduced in Sec. \ref{sec. data} and also assume the same LSST galaxy sample for cleaning. In some cases, where we wish to include high redshift BAO in our forecasts, we include futuristic BAO measurements from spectroscopic surveys achievable in the next decade with Megamapper \cite{Megamapper}. We do not use Ly-$\alpha$ BAO from DESI as these do not significantly improve the constraints compared to the lensing-only case at high redshifts. Since these surveys can measure structure on relevant scales at redshifts of $z>2$ with $\text{SNR}>1$, we approximate the corresponding errors of $d_A(z)$ and $H(z)$ by simply scaling the results forecast for DESI \cite{Font-Ribera2014}  by the volume surveyed by Megamapper at each redshift bin to estimate the required $f_k=r_s/d_V$  uncertainties.\\

\subsubsection{Fisher matrix analysis}\label{LR}

We consider a set-up where $\sum{m}_\nu$ is allowed to vary along with the other $\Lambda$CDM parameters in a flat universe with $K=0$. We write the parameters in the vector:  \begin{equation}
    \overrightarrow{\theta}=\{\theta_{MC},\Omega_bh^2,\Omega_ch^2,\tau,n_s,A_s,\sum{m}_\nu\}
\end{equation}

The information from the primary CMB is included by computing the Fisher matrix $F^{\text{CMB}}$ as
\begin{widetext}
\begin{equation}
  F^{\text{CMB}}_{ij}=\sum_\ell\frac{2\ell+1}{2}f_{\text{sky}}\Tr\Big[\frac{\partial{\mathbb{C}^{\text{CMB}}_\ell}}{\partial\theta^i}(\mathbb{C}^{\text{CMB}}_\ell)^{-1}\frac{\partial{\mathbb{C}^{\text{CMB}}_\ell}}{\partial\theta^j}(\mathbb{C}^{\text{CMB}}_\ell)^{-1}\Big]
 \end{equation}
\end{widetext}
where $\mathbb{C}^{\text{CMB}}_\ell$ is the covariance matrix of the CMB 
\begin{equation}
\mathbb{C}^{\text{CMB}}_\ell=
\begin{pmatrix}C^{\text{TT}}_\ell & C^{\text{TE}}_\ell\\
C^{\text{TE}}_\ell & C^{\text{EE}}_\ell
\end{pmatrix}
\end{equation}
Here, $C^{\text{TT}}_\ell$ is the power spectrum of the temperature anisotropies including noise, $C^{\text{EE}}_\ell$ is the power spectrum of the E-mode anisotropies, and $C^{\text{TE}}_\ell$ is their cross-spectrum. We will use the unlensed primary power spectra here to avoid overcounting the lensing information which will come from the lensing Fisher matrix $F^{\kappa\kappa}$ (our estimate is therefore slightly conservative for lensing-derived parameters), given explicitly as
\begin{equation}
    F^{\kappa\kappa}_{ij}=\sum_{L}\frac{\partial C^{\kappa\kappa}_L}{\partial\theta_i}{{\mathbb{C}_L^{\kappa\kappa}}}^{-1}\frac{\partial C^{\kappa\kappa}_L}{\partial\theta_j}
\end{equation}
where the lensing covariance matrix ${\mathbb{C}_L^{\kappa\kappa}}=2\times({C}_L^{\kappa\kappa}+N^{\kappa\kappa}_L)^2/f_{\text{sky}}/(2L+1)$ is diagonal and contains lensing reconstruction noise; whenever the cleaned lensing field is used, we simply replace the lensing spectra by the LSST-galaxy-cleaned versions. In addition, we include the effective galaxy shot-noise $N^{XX}_L=\sum_ic^2_iN^{gg}_i$, obtained by weighting using Eq. \ref{weights} the shot-noise $N^{gg}_i$ of each galaxy bin as part of the covariance matrix 

\begin{equation}
   {\mathbb{C}_L^{\kappa\kappa}}=\frac{2}{f_{\text{sky}}(2L+1)}\times\Bigg[{C}_L^{\kappa_\text{clean}\kappa_\text{clean}}+N^{\kappa\kappa}_L+N^{XX}_L\Bigg]^2
\end{equation}

so that the Fisher matrix for the cleaned lensing field is given by
\begin{equation}
    F^{\kappa_\text{clean}\kappa_\text{clean}}_{ij}=\sum_{L}\frac{\partial C^{\kappa_\text{clean}\kappa_\text{clean}}_L}{\partial\theta_i}{{\mathbb{C}_L^{\kappa\kappa}}}^{-1}\frac{\partial C^{\kappa_\text{clean}\kappa_\text{clean}}_L}{\partial\theta_j}
\end{equation}

The above setup also ignores the cross power spectra of CMB lensing with CMB temperature and E mode polarization, as these are only non-zero on very large scales via late-time effects on the CMB such as the Integrated Sachs-Wolfe effect, which are unlikely to significantly bias neutrino mass constraints. The combined set-up of the 2-point and 4-point functions is consistent with the CMB-S4 experiment as laid out in  Sec. \ref{lens spec} above.

We include a Gaussian prior with width  $\sigma(\tau)=0.005$ for  $\tau$, the optical depth from reionization. Assuming that these datasets are independent, the total Fisher matrix used is thus 

\begin{equation}\label{fisher}
    F_{ij}={F_{ij}}^{\text{CMB}}+{F_{ij}}^{\kappa\kappa}+{C^{-1}_{\text{prior}}}_{ij}
\end{equation}

where $C_{\text{prior}}$ is the sum of the $\tau$ and, where applicable,  BAO prior obtained from Megamapper. BAO is sensitive to the sum of the neutrino and CDM density, hence its inclusion allows for reducing the errors on $\sum{m}_\nu$ by breaking the degeneracy between CDM and  $\sum{m}_\nu$ of the CMB data.

The $1\sigma$ errors on the parameters $i$, marginalized over the other parameters, are given by

\begin{equation}
    \sigma_i=\sqrt{(F^{-1})_{ii}}
\end{equation}

\subsection{Bias induced in the inference of the neutrino mass}

Primary CMB measurements and lensing from Planck combined with BAO have placed constraints on the total neutrino mass of $\sum{m}_\nu\leq0.12\si{eV}$\cite{Aghanim2020} at the $95\%$ level. Using the above Fisher formalism,  combining CMB-S4 lensing and primary CMB as well as large-scale structure observations from the full DESI BAO would place tighter bounds on the neutrino mass sum with constraints of $\sim 20 \si{meV}$. However, as discussed already in \cite{Allison2015,Mishra-Sharma2018}, the constraints of $\sim 20\si{meV}$ are derived in the framework of the $\Lambda$CDM model and are hence, to some extent, model-dependent; extending the model by including effects like dark energy and curvature, which are degenerate with the effects produced by massive neutrinos, can degrade the constraints on neutrino mass significantly. These extensions are often required for robust measurement of $\sum{m}_\nu$ because the true lensing power spectrum could include unknown effects induced by non-standard dark energy at low redshifts, which differs from the lensing power spectrum computed assuming $\Lambda\text{CDM}$. Such deviations might mimic the neutrino mass signature leading to a biased inference of $\sum{m}_\nu$.

Rather than extending the $\Lambda$CDM model, we produce forecasts of constraints on $\sum{m}_\nu$ assuming $\Lambda$CDM, but with a lensing map that only contains high redshift information. This approach has the advantage of being agnostic about the dark energy model -- assuming only that dark energy is negligible at sufficiently high redshifts -- and is simpler to implement than the alternative of taking into account all the possible dark energy scenarios in the model. The intuition behind our method is that since (we assume) the effect of dark energy becomes dominant only at low redshifts, cleaning the low redshift contribution of the lensing maps removes the degeneracy caused by dark energy, leading to an unbiased measurement of $\sum{m}_\nu$. Assuming Gaussian errors, we can quantify the bias $B_i$ induced in the inference of a parameter $\theta_i$ using the Fisher formalism as \cite{Knox_1998re,Linder_2006,2008amara,Huterer_2007}\footnote{Note that Ref.~\cite{2008amara} assumes that the data is biased. On the other hand, this paper assumes that the theoretical model to be fitted to data is biased. Thus, we compute the covariance and the Fisher matrix in the equation with biased theoretical spectra. More explicitly, the theoretic spectra used here assume the $\Lambda$CDM model while the true model describing the data is given by an  extension of $\Lambda$CDM with dark energy parametrised by $w_0$ and $w_a$. }

\begin{equation}
    B_i=F^{-1}_{ij}\sum_\ell\Delta{C_\ell}\text{Cov}^{-1}[C_\ell,C_\ell]\frac{\partial{C_\ell}}{\partial\theta_j}
\end{equation}\\

Here $F_{ij}$ is the Fisher matrix of Eq. \ref{fisher} obtained from the power spectra assuming $\Lambda\text{CDM}$ parameters and $\Delta{C^{\kappa\kappa}_L}$ is the change in the lensing power spectrum due to dark energy effects.

\begin{equation}
   \Delta{C^{\kappa\kappa}_L}\equiv{C^{\kappa\kappa}_L}\mid_{\text{DE}}-{C^{\kappa\kappa}_L}\mid_{\Lambda\text{CDM}}.
\end{equation}

Here we are making the approximation that the CMB power spectra are unaffected by the change in dark energy at fixed $\theta_{\mathrm{MC}}$, since keeping the sound horizon angle constant means that the power spectrum can only be minimally affected by late-time physics. This assumption was numerically verified using CAMB, where only per cent-level deviations are observed in the 2-point function at very low multipoles at $\ell<50$. Deviations are otherwise negligible at higher multipoles; we note that for our forecast we restrict CMB-S4 primary CMB multipoles to larger than $\ell > 100$.

Including the effect that dark energy has on lensing and BAO, the bias expression becomes

\begin{widetext}
\begin{equation}
    B_i=F^{-1}_{ij}\Big[\sum_L\frac{2L+1}{2}f_{\text{sky}}\frac{\partial{C^{\kappa\kappa}_L}}{\partial\theta_j}\frac{\Delta{C^{\kappa\kappa}_\ell}}{(C^{\kappa\kappa}_L+N^{\kappa\kappa}_L)^2}+\Delta{f}_k\frac{1}{\sigma^2_k}\frac{\partial{f_k}}{\partial{\theta}_j}\Big]
\end{equation}
\end{widetext}

To test our cleaning procedure, we will consider that, although our analysis is performed assuming standard $\Lambda$CDM, the true cosmological model is well-described by a simple model of dynamical dark energy using the standard Taylor expansion in the scale factor
\begin{equation}
    w(a)=w_0+w_a(1-a).
\end{equation}
Despite this example, we emphasize that we expect our cleaning method to be generally applicable, regardless of the details of the dark energy model. Indeed we will also test that the method works well with dark energy models with arbitrary $w=w(z)$ dependence.

\subsection{Results}\label{sec:res}

\begin{figure*}
\includegraphics[width=\textwidth]{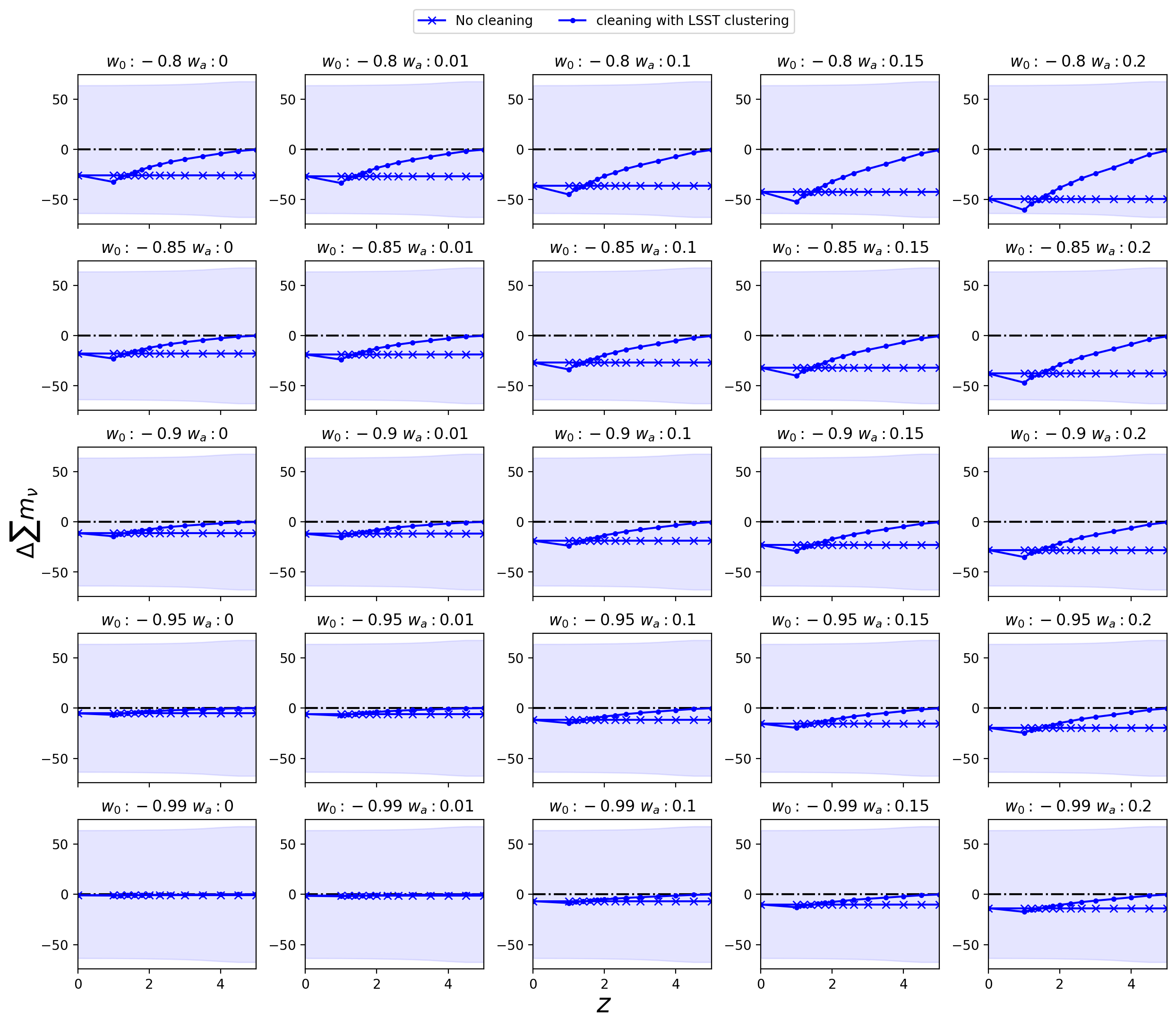}
\caption{\label{fig:nohubble} Bias induced in the neutrino mass sum $\Delta{\sum{m}_\nu}$ measurement due to incorrectly assuming $\Lambda$CDM when the true cosmology is described certain $w_0-w_a$ dark energy model. The bias in meV is shown as a function of the maximum redshift to which we have cleaned the lensing map using the LSST galaxies. A conservative scale cut of $k_\text{cut}=0.3h\si{Mpc}^{-1}$ and lensing multipole $L_\text{max}=200$ is assumed. It can be seen that by cleaning the low-$z$ contributions from the lensing map to a sufficiently high redshift, unbiased measurements of the neutrino mass can even be obtained if we assume the wrong dark energy model. 1$\sigma$ measurement errors are indicated with the blue shaded region. }
\end{figure*}

Fig. \ref{fig:nohubble} shows the biases on neutrino mass due to a range of non-standard dark energy models. The dashed lines represent the base cases in which no cleaning is applied, whereas the solid lines correspond to the bias when our cleaning procedure is applied using galaxy bins up to a certain $z$ only, following the prescription laid in Sec. \ref{Sec. Cleaning} above. It can be seen that in agreement with our intuition in Sec. \ref{sec. weight}, including all the LSST galaxy bins up to redshift $5$ effectively reduces the neutrino mass sum bias to close to zero, with only minimal increase in the statistical error shown as the green bands. Even in an extreme case where we have assumed $\Lambda$CDM but the `true' dark energy model is  $w_0=-0.8$ and $w_a=0.20$, removing the $z<5$ portion of the lensing map reduces the bias from $-48\si{meV}$ to zero. The slight increase in bias observed around $z=1.5$ can be explained by the fact that the effect of dark energy on the lensing spectra is not a monotonic function with respect to the amount of low redshift removed. Hence removing only the very low redshifts can still exacerbate the difference between the true cosmology and $\Lambda$CDM compared to the case where no cleaning is applied; the reduction in bias only starts to occur when all the low redshifts containing significant dark energy contributions are removed.

\begin{figure*}
\includegraphics[width=\textwidth]{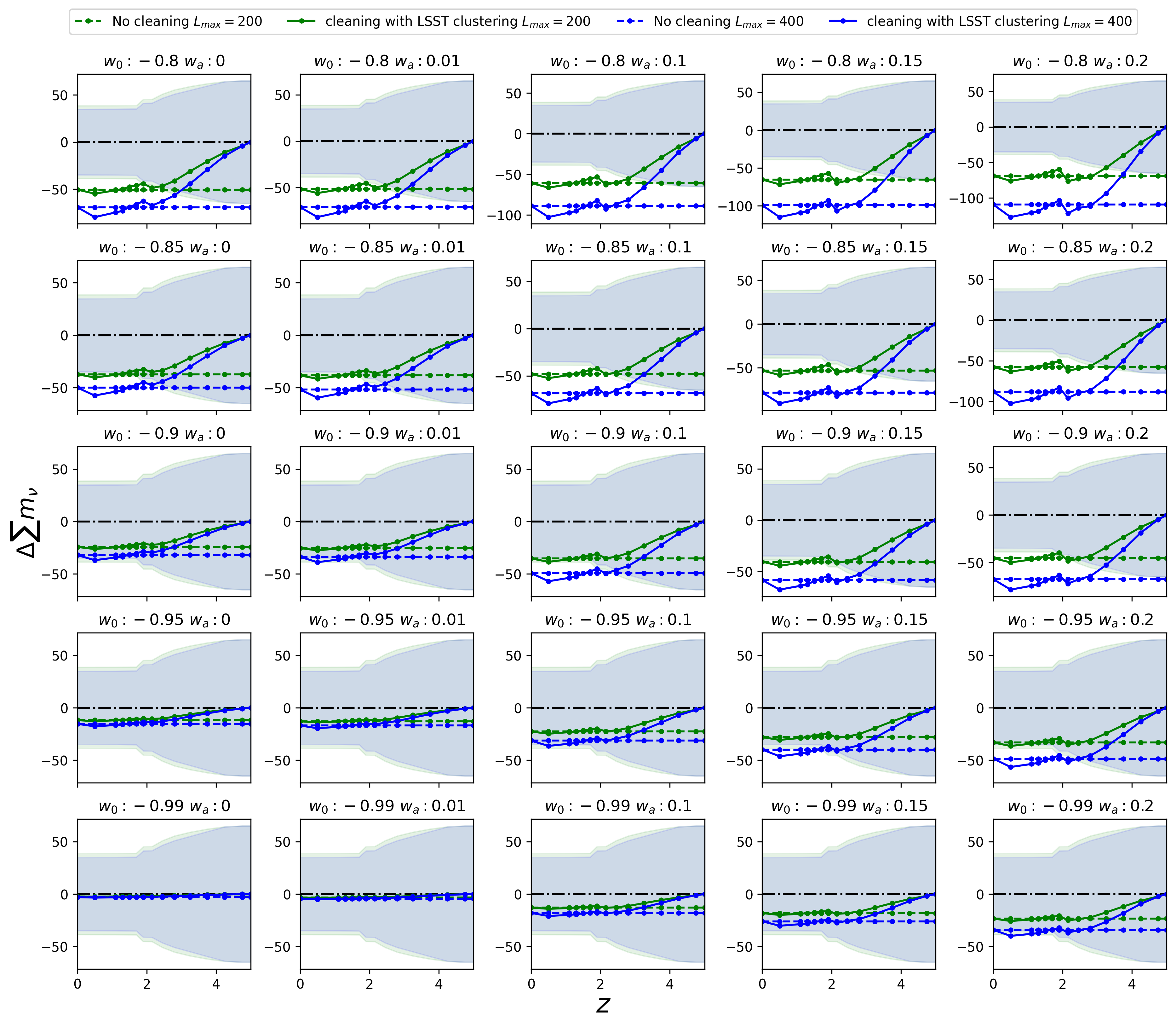}
\caption{\label{Fig.mega} As in Fig.~7, but now also including a futuristic BAO survey such as Megamapper (constraints assuming $L_\text{max}$ of 200 and 400 are shown in green and blue). It can again be seen that by cleaning the low-$z$ contributions from the lensing map to a sufficiently high redshift, unbiased measurements of the neutrino mass can even be obtained if we assume the wrong dark energy model. The slight increase in bias at around $z=2$ is produced by the introduction of the high redshift BAO, which also incurs a negative bias in the determination of neutrino mass sum. }
\end{figure*}

The lensing multipoles $L$ used in the above forecast range from 20 to 200. The $L_\text{max}$ multipole is set by the scale cut of the lowest galaxy bin (for the conservative forecast, we use $k_\text{cut}=0.3h\si{Mpc}^{-1}$) which corresponds to $L_\text{max}=k_{\text{cut}}\chi(z_{g_1})\sim200$ since angular scales above this $L_{\text{max}}$ contain low redshift lensing signal  which is not removed by the galaxies in the first redshift bin. The above forecast, which only contains CMB-S4 primary CMB and $z>5$ lensing will provide neutrino measurements with  $\sigma(\sum{m}_\nu)=62\si{meV}$. This is comparable to the forecast obtained using CMB-S4 primary CMB and lensing alone of $\sigma(\sum{m}_\nu)=53\si{meV}$\cite{Allison2015} (when including $L_{\text{max}}=3000$), but with the main difference that our method should be nearly model-independent. Our goal of obtaining a nearly model-independent constraint on $\sum{m}_\nu$ also limits our ability to exploit the full constraining power of BAO surveys (since current high resolution BAO are mostly at low redshifts which could be affected by dark energy modelling); however, using a futuristic high redshift BAO measurement from Megamapper will allow neutrino constraints of order $\sigma(\sum{m}_\nu)=39\si{meV}$ ($\sigma(\sum{m}_\nu)=34\si{meV}$ if using $L_\text{max}=400$, corresponding to a scale cut of $k_\text{cut}=0.6h\si{Mpc}^{-1}$)  as shown in Fig. \ref{Fig.mega}. Here we only include the BAO at redshifts equal or greater to the redshifts over which we use the LSST galaxies, which explains the broadening of the errors at higher redshifts. 

\begin{figure}
\includegraphics[width=\linewidth]{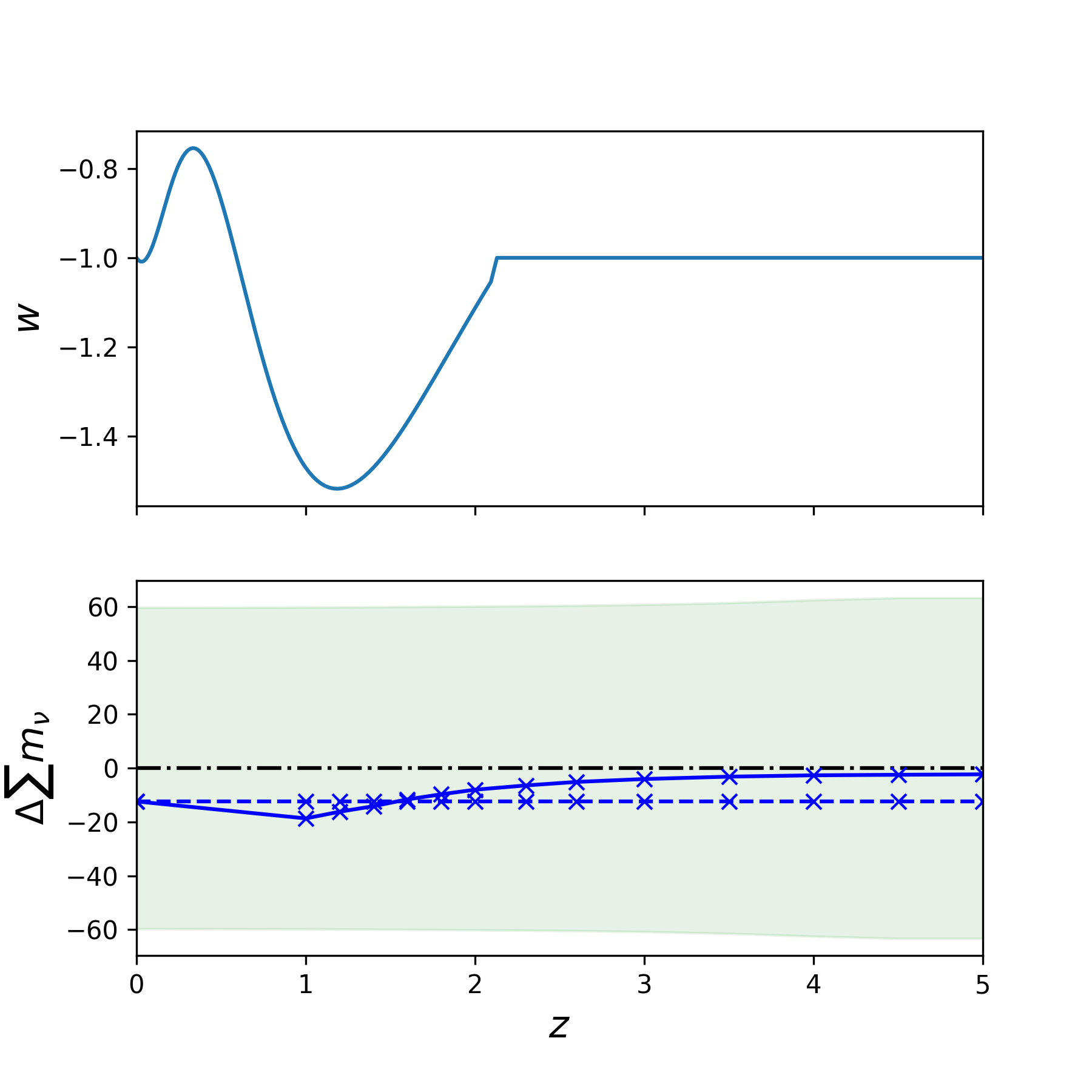}
\caption{\label{fig:wflexible} \textit{Top panel}: Example of a (toy model) dark energy equation of state evolution which does not follow the conventional $w_0-w_a$ parametrization. \textit{Bottom panel}:  As in Fig.~7, the bias induced in the neutrino mass sum $\Delta{\sum{m}_\nu}$ as a function of the maximum redshift cleaned with LSST galaxies. The results show that even with a highly non-standard dark energy evolution at low redshifts, unbiased results on neutrino mass can be obtained with our cleaning procedure.}
\end{figure}

Fig. \ref{fig:wflexible} illustrates that this cleaning method is flexible and can also work when the true cosmology has dark energy more complicated than that given by the $w_0-w_a$ parametrization. The dark energy equation of state in the upper panel, which is a toy example that oscillates with redshift, is not described using the $w_0-w_a$ framework, but the bias on the neutrino mass is still suppressed when using cleaned lensing maps.

\section{Discussion and Conclusions}\label{sec.discussion}
In this work, we discuss the production and application of a lensing map containing only high redshift information. A high-$z$ mass map can be constructed in a model-insensitive manner by subtracting suitably scaled correlated tracers, such as galaxy density maps, from the lensing field. Our lensing cleaning method is similar to that of delensing, with the critical difference that instead of minimizing the variance of the resultant delensed field including noise, we minimise only the lensing signal at low $z$.

The resulting high-$z$ lensing map allows us to probe the growth of structure at redshifts which are currently not accessible to other cosmological probes. With CMB-S4 lensing and Rubin--LSST galaxies (and a scale cut of $k=0.3h\si{Mpc}^{-1}$, which corresponds to a usable $L_\text{max}$ of 200), we forecast measurements of the amplitude of structure  at high redshifts $z>3.75$ and $z>5$ to within $2.3\%$ and $3.3\%$ respectively. We explore the impact of uncertainties in the lensing-galaxy cross spectrum in determining the cleaned lensing map and find that these uncertainties do not greatly affect the results.

The high redshifts probed by the cleaned lensing maps also correspond to periods when dark energy, modification to general relativity and non-linear effects due to baryonic physics are often assumed to be subdominant or at least less important than at lower redshifts. As an example application of this, we note that a high-$z$ mass map could provide a clean and arguably less model-dependent constraint on the neutrino mass sum; we forecast that a high-$z$ mass map constructed from CMB-S4 and Rubin-LSST galaxies can provide a competitive neutrino mass determination ($\sigma(\sum{m}_\nu)=62\si{meV}$ or $39\si{meV}$ including future high-$z$ BAO), and show that this measurement has only negligible sensitivity to the presence of non-standard dark energy models. Furthermore, the same cleaning method also removes potential degeneracies in the determination of neutrino mass sum with effects induced by many modified gravity models, as discussed in detail in Appendix \ref{appendixA}, and may reduce sensitivity to baryonic feedback (as in \cite{McCarthy2020}) and matter non-linearities.


There are other potential applications of variants of our CMB lensing redshift-cleaning method. In Appendix \ref{HR}, we discuss cleaning certain redshifts to improve the signal-to-noise-ratio of lensing-galaxy cross-correlations. In particular, we find that removing the high redshift information from a lensing map using the CIB significantly improves the determination of parameters at low-$z$, such as linear bias and growth factor, by $40-60\%$ for current and upcoming CMB surveys. Our redshift-cleaning technique might also be useful for mitigating the intrinsic alignment contamination to the cross-correlation between CMB lensing and cosmic shear, if contributions to CMB lensing from source galaxy redshifts can be removed. We hope to explore these and other applications of redshift-cleaning methods and their implementation in data in future work.

\begin{acknowledgments}
We thank Fiona McCarthy and  Ant\'{o}n Baleato Lizancos for useful discussions and feedback. FQ acknowledges the support from a Cambridge Trust scholarship. BDS acknowledges support from the European Research Council (ERC) under the European Union's Horizon 2020 research and innovation programme (Grant agreement No. 851274) and from an STFC Ernest Rutherford Fellowship. TN acknowledges support from JSPS KAKENHI Grant No. JP20H05859 and No. JP22K03682. OD acknowledges support from SNSF Eccellenza Professorial Fellowship (No. 186879). This research used resources of the National Energy Research Scientific Computing Center (NERSC), a DOE Office of Science User Facility supported by the Office of Science of the U.S. Department of Energy under Contract No. DE-AC02-05CH11231.
\end{acknowledgments}

\appendix

\section{High redshift lensing map and its sensitivity to modified gravity }\label{appendixA}
An additional advantage of our low-$z$ cleaning procedure is that it can make constraints from the cleaned lensing field insensitive to modified gravity effects. This is especially relevant at low redshifts where non-standard effects are particularly well motivated due to the still poorly understood cosmic acceleration \cite{Bertschinger_2008,2012mod}. Theories that propose deviations from general relativity can have both different background evolution equations and different growth of structure on scales where the neutrino mass suppression occurs. We will argue in the following paragraphs that any degeneracy of neutrino mass with low-$z$ modified gravity effects can similarly be broken when the low-$z$ information about the growth of structures is removed, as described previously. 

Modified theories of gravity differ from General Relativity in two primary ways (e.g. \cite{Zhang2007}). 
First, they change the relationship between the two scalar potentials of the perturbed FLRW metric. With a metric in the conformal Newtonian gauge for scalar perturbations, ${\rm d}s^2=(1+2\psi){\rm d}t^2-a^2(1-2\phi){\rm d}\bm{x}^2$, the relation $\psi=\phi$ is no longer true even in the absence of anisotropic stress. 
Second, the relation between the potential and density perturbations, as related by the Poisson equation, is also modified. In the following discussion, we consider modified gravity theories that can be parametrized by the following two equations \cite{Simpson:2012:modified-gravity}:\footnote{ \url{https://wwwmpa.mpa-garching.mpg.de/~komatsu/lecturenotes/Alex_Barreira_on_Modified_Gravity.pdf}}
\begin{align} 
    \nabla^2\psi(\chi,\bm{x}) &= 4\pi G\bar{\rho}_{\rm m}[1+\mu(\chi)] \delta(\chi,\bm{x})
    \,, \label{eq.poisson}
    \quad \\ 
    \Psi(\chi,\bm{x}) &= [1+\Sigma(\chi)]\Psi_{\rm GR}(\chi,\bm{x}) 
    \,, \label{eq.weyl} 
\end{align}
where $\Psi\equiv (\psi+\phi)/2$ is the so-called Weyl potential \cite{LEWIS_2006} entering into the lensing potential and the subscript ``GR'' indicates the quantity in general relativity. The functions $\mu$ and $\Sigma$ are exactly zero in the standard $\Lambda${CDM} model. Here, we ignore the scale dependence of $\mu$ and $\Sigma$, although it is straightforward to apply the following discussion to the case where $\mu$ and $\Sigma$ vary slowly with scale.

Now we calculate the lensing mass field $\widetilde{\kappa}$ from the modified Weyl potential \eqref{eq.weyl} above. The first term of \eqref{eq.weyl} results in the lensing convergence in general relativity, $\kappa$, while the second term changes the lensing convergence by:
\begin{equation}
    \Delta\kappa(\hat{\bm{n}}) = \tilde{\nabla}^2 \int^{\chi_*}_0 {\rm d}\chi \frac{H(z)W^\kappa(z)}{\chi^2}\Sigma(\chi)\nabla^{-2}\delta(\chi,\chi\hat{\bm{n}}) 
    \,,
\end{equation}
where $\tilde{\nabla}$ is the covariant derivative on the unit sphere. We assume that $\Sigma$ varies slowly with $\chi$ so that we can pull $\Sigma$ out of the integral. Furthermore, we assume that the modification is only important at low redshifts. Then, the lensing convergence above could be written as
\begin{equation}
    \Delta\kappa_{\bm{L}} \simeq \Sigma \kappa_{\text{low},\bm{L}}
    \,,
\end{equation}
where we have moved $\Sigma$ outside of the integral. Under modified gravity, and in the notation of the draft, the CMB lensing field is therefore affected as:
\begin{equation}
   \widetilde{\kappa} = \kappa_{\text{high}}+\kappa_{\text{low}}(1+\Sigma) 
   \label{Eq:modified-kappa}
\end{equation}
On the other hand, a galaxy tracer will be affected by modified gravity through the Poisson equation and the growth of the density perturbations changes \cite{Simpson:2012:modified-gravity}.  Assuming that the modification changes the growth of the linear perturbation by an overall amplitude, we can write the galaxy tracer at low $z$ as $\tilde{X}=(1+D)X$ where $X$ represents the case of standard cosmology. 

If the galaxy tracer is originally correlated with the low-$z$ lensing field by $T_L$ in general relativity, the galaxy tracer is given by: 
\begin{equation}
    \tilde{X}_{\bm{L}} = T_L (1+D) \kappa_{\text{low},\bm{L}}
    \,. 
\end{equation}
If we require that the weights for cleaning are consistent with the measured spectra, our cleaning procedure will be immune to these scaling functions. To see this, we note that the observed spectra, $C_L^{\kappa X}$ and $C_L^{XX}$, are rescaled, respectively, by $(1+\Sigma)(1+D)$ and $(1+D)^2$. The observed galaxy tracer, $\hat{X}_{\bm{L}}$, is also rescaled by $(1+D)$, the second term in Eq.~\eqref{eq. kk} becomes $(1+\Sigma)\kappa_{\rm low}$ which exactly cancels with the modified low-$z$ lensing convergence involved in the first term of Eq.~\eqref{eq. kk} (i.e. the second term in Eq.~\eqref{Eq:modified-kappa}). 

\section{Selective delensing in the context of cross-correlations}\label{HR}

We explore the prospects of using the lensing cleaning method discussed above in applications related to cross-correlations. The lensing kernel contains information about the integrated mass distribution up to the last-scattering surface. However, most photometric and spectrometer galaxy surveys are restricted to lower redshifts. Taking the cross-correlation of the lensing field with late time galaxy probes $C_L^{\kappa{g}}$ helps constrain parameters such as the growth of structure and galaxy bias at late times. Assuming Gaussian errors, the expected error in these cross-correlation spectra is given by

\begin{equation}
    \Delta{C_L^{\kappa{g}}}=\sqrt{\frac{1}{(2L+1)f_{\text{sky}}}}[{C^{\hat{\kappa}\hat{\kappa}}_LC^{\hat{g}\hat{g}}_L}+(C^{\hat{\kappa}\hat{g}}_L)^2],
\end{equation}

The linear bias $b_1$ which we assume to be scale and redshift independent can be obtained from the following relationship: $b_1\propto{C^{gg}_L/C^{\kappa{g}}_L}$ with an error proportional to $\Delta{C_L^{\kappa{g}}}$ if we ignore the error in determining $C^{gg}_L$ which is usually smaller than the cross-correlation error of $C^{\kappa{g}}_L$. We can see that if we could reduce the cosmic variance due to $C^{\kappa\kappa}_L$, this could lead to smaller errors in the cross-correlation error and subsequently a smaller error in the linear bias. To do so requires increasing the correlation between the lensing and the galaxy field since the high redshift portion of the lensing field has little correlation with the galaxy field, and its presence only increases the cross-correlation error. 

Similar to the case of the neutrino application, here we can form a low redshift only lensing map using the CIB as a proxy of lensing at high redshifts.

\begin{equation}\label{eq. kkhigh}
    \hat{\kappa}^{\text{clean}}_\vec{L}=\hat{\kappa}_\vec{L}-\frac{C^{\hat{\kappa}{\hat{I}}}_L}{C^{{I}{I}}_L+{N^{II}_L}}\hat{I}_{\vec{L}},
\end{equation}

where $\hat{\kappa}_\vec{L}$ is the original CMB lensing convergence. We note that, since our goal, in this case, is here to minimize the low-redshift variance, the weights must include noise, similar to the standard delensing case.\\

\subsection{Modelling the lensing-CIB correlation}\label{sec:cib}

Since the goal here is to null the high redshift contribution to the lensing map, we need to take a highly correlated tracer with CMB lensing and subtract that from it. An ideal candidate for such as tracer is the cosmic infrared background (CIB) \cite{cib} which is strongly correlated with the CMB lensing potential due to their extensive overlap of the redshift kernels of both fields \cite{Sherwin2015}. The resulting cleaned lensing spectrum can be obtained by estimating the correlation coefficient of the CIB with lensing $\rho_\ell$, and it is given by $C^{\kappa\kappa}_L\rightarrow{C^{\kappa\kappa}_L(1-\rho^2_L})$.\\

To model the CIB, we adopt the single energy distribution (SED) model of \cite{2010hall} with the following kernel:
\begin{equation}\label{eq.cib}
    W^{\text{CIB}}_L(\chi)=b_c\frac{\chi^2}{(1+z)^2}\exp{\Big[-\frac{(z-z_c)^2}{2\sigma^2_z}\Big]}f_{\nu(1+z)},
\end{equation}
where $b_c$ is the normalization obtained by matching the $C^{\text{CIB}\times\kappa}_l$ spectrum with the empirical spectrum obtained from the Planck $545\si{GHz}$ channel and $z_c=\sigma_z=2$ describes the redshift distribution of the CIB intensity. $f_\nu$ describes the SED of the CIB source
\[
    f_\nu= 
\begin{cases}
    \Big(e^{\frac{h\nu}{kT}}-1\Big)^{-1}\nu^{\beta+3},& \text{if } \nu\leq\nu^\prime\\
    \Big(e^{\frac{h\nu^\prime}{kT}}-1\Big)^{-1}{\nu^{\prime}}^{\beta+3}\Big(\frac{\nu}{\nu^\prime}\Big)^{-\alpha},&  \nu>\nu^\prime
\end{cases}
\]
with $T=34\si{K}$ , $\alpha=\beta=2$ and the power law transition occurring at $\nu^\prime\approx4955\si{GHz}$. \\

\begin{figure}
\includegraphics[width=\linewidth]{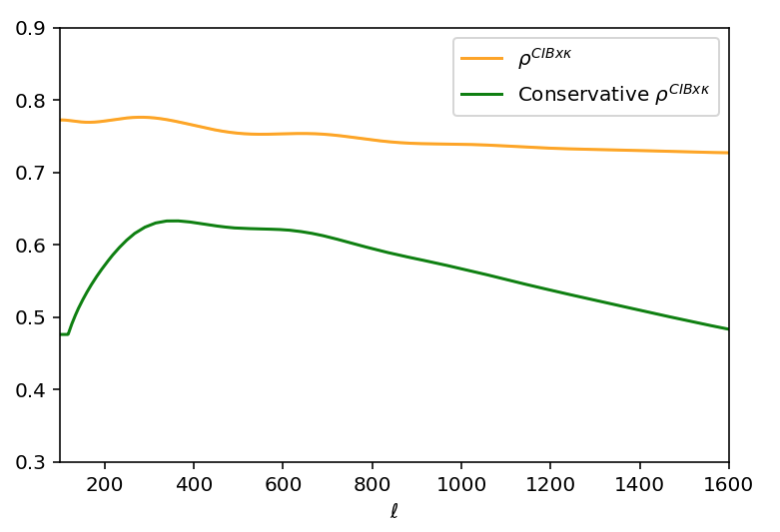}
\caption{\label{fig:cib_corr} Correlation coefficients of the CIB with CMB lensing. The orange curve is obtained from the SED model, with parameters fitted to match the Planck CIB-lensing cross-correlation spectrum. The green curve is the correlation obtained from the GNILC CIB map.}
\end{figure}

 The correlation coefficient for this CIB with lensing in orange is shown in Fig. \ref{fig:cib_corr}, where we find a high degree of correlation reaching nearly 0.8 for a large range of multipoles illustrating that the CIB is an excellent tracer of CMB lensing. Shown in green is a more conservative correlation coefficient which we also test in the cleaning procedure. The correlation is obtained from maps of CIB constructed via the Generalized Needlet Internal Linear Combination (GNILC) algorithm) at 353 GHz with shot noise and contribution from galactic dust emission included \cite{2016cib}. Large Galactic dust residuals contaminate the large scales of this CIB auto and cross spectra, and hence only scales $L\geq100$ are included.\\

\subsection{Forecasting method}
\subsubsection{Angular Power Spectra}

To illustrate the above intuition, we use a toy galaxy field, the CIB field (See \ref{sec:cib} for details about its modelling) and the lensing convergence field to construct the 2-point angular power spectra: $C^{\kappa\kappa}_L,C^{\kappa{g}}_L,C^{\kappa{\text{CIB}}}_L$ ,$C^{\kappa\text{CIB}}_l=L$ and $C^{\text{CIB}\text{CIB}}_L$. In the Limber approximation, the angular power spectra are given by

\begin{equation}
    C^{\alpha\beta}_\ell=\int\frac{dzH(z)}{\chi^2(z)}W^\alpha(z)W^\beta(z)P\Big(k=\frac{l+\frac{1}{2}}{\chi(z)},z\Big),
\end{equation}

where $\alpha,\beta\in(\kappa,g,\text{CIB})$.

The CIB window function is shown in Eq. \ref{eq.cib}. For the galaxy field, we consider a Gaussian galaxy density field with mean redshift  $z_0=0.1$ and width $\sigma_z=2$ with a galaxy density of $\bar{n}=1.06\si{arcmin}^2$ and window function $W_g$ given by

\begin{equation}
    W^g=\frac{b_1A_z}{\sqrt{2\pi\sigma^2}}e^{\frac{-(z-z_0)^2}{2\sigma^2}}.
\end{equation}

Here $b_1$ is the linear bias with a fiducial value of $b_1=2$, $A_z$ is a redshift dependent amplitude which is degenerate with $b_1$ and in this example we take as values $A_{\text{low}}=1.2$ for $z<0.5$ and $A_\text{high}=1.2$ for $z>0.5$. This factor is applied to the CIB and lensing kernels as well.

\subsubsection{Fisher Analysis}

For the CIB delensing application, we also consider lensing experiments with higher noise levels up to $10\si{\mu{K}}'$ attainable by a Simons Observatory-like experiment with the specifications:
beam FWHM = $1.4'$, $\Delta_T=8\si{\mu{K}}'$, $\Delta_P=11.3\si{\mu{K}}'$ and $f_{\text{sky}}=0.3$ and lensing experiments attained with current ground-based observations such as AdvACT \cite{2016act} with beam FWHM = $1.4'$, $\Delta_T=15\si{\mu{K}}'$, $\Delta_P=15.3\si{\mu{K}}'$ and $f_{\text{sky}}=0.3$. The minimum variance reconstruction noise by combining the different temperature and polarization channels for the different experiments is shown in Fig. \ref{fig:lensing}. 

The observables we consider are the lensing-galaxy cross spectrum $C^{\kappa{g}}_L$, the galaxy auto spectrum  $C^{g{g}}_L$ and the CMB lensing auto spectrum, $C^{\kappa\kappa}_L$ or the high redshift cleaned version $C^{\kappa_c\kappa_c}_L$ which is related to the uncleaned one via the CIB-lensing correlation $\rho_L$, $C^{\kappa_c\kappa_c}_L=C^{\kappa\kappa}_L(1-\rho^2_L)$. In the above example, since the overlap between the galaxy field and the CIB field is very small, we assume that $C^{\kappa_cg}_L\approx{C}^{\kappa{g}}_L$.  We use a Gaussian covariance matrix $\text{Cov}^{\alpha_1\beta_1,\alpha_2\beta_2}_{L_a,L_b}$  where $\alpha_{1,2},\beta_{1,2} \in (\kappa,g)$. For the CMB lensing convergence and the auto and cross power spectra entering the covariance matrix we take into account the lensing reconstruction noise $N^{\kappa\kappa}_L$  and the shot noise $N^{gg}_L=1/\bar{n}$. \\

We construct the Fisher matrix as

\begin{equation}
    F_{ij}=\sum_{\substack{\alpha_1\beta_1,\\ \alpha_2\beta_2}}\sum_L\frac{\partial{C^{\alpha_1\beta_1}_L}}{\partial\theta_i}[\text{Cov}^{\alpha_1\beta_1,\alpha_2\beta_2}_{L_a,L_b}]^{-1}\frac{\partial{C^{\alpha_2\beta_2}_L}}{\partial\theta_j}
\end{equation}

where $\vec{\theta}=\{b_1,A_{\text{low}},A_{\text{high}}\}$ are the parameters that we vary.

\subsection{Results and interpretation}

\begin{figure}
\includegraphics[width=\linewidth]{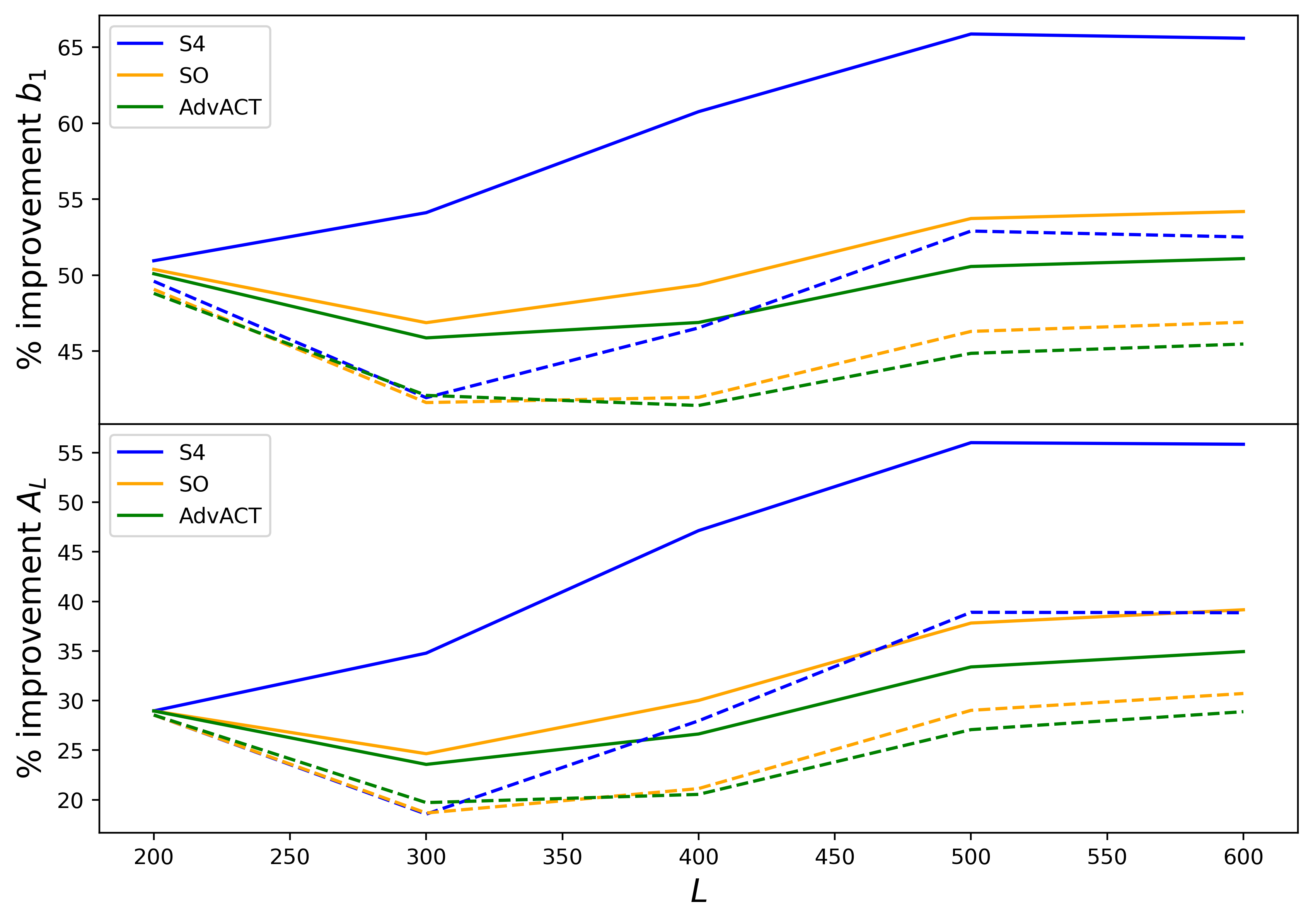}
\caption{\label{fig:lowcib} \textit{Top panel}: Reduction in the forecasted errors on the linear bias. \textit{Bottom panel}: Similar improvements were observed in the constraints on the amplitude of structure. The forecast is performed for the cross-correlation between a low-$z$  Gaussian galaxy sample and the lensing field cleaned with CIB. The solid line uses the SED CIB model, and the dotted lines use the CIB that has a more conservative correlation coefficient with CMB lensing, see Fig. \ref{fig:cib_corr}}
\end{figure}

Fig \ref{fig:lowcib} shows that for a CMB-S4 CMB experiment, our method of removing the high redshift content of the lensing map can lead to $60\%$ improvement in the determination of the linear bias and $50\%$ improvement in a measurement of the amplitude of structure. The main factor in tightening the constraints still comes from the degree of correlation between the CIB and lensing. This improvement is degraded when using the CIB field with a smaller degree of correlation. For those cases, the CMB noise levels are not that important, with experiments like SO and AdvACT achieving similar improvements of $20\%-30\%$.

\nocite{*}

\bibliography{apssamp}

\end{document}